\documentclass[a4paper,fleqn,usenatbib]{mnras}

\usepackage[T1]{fontenc}
\usepackage{ae,aecompl}

\usepackage{graphicx}	% Including figure files
\usepackage{amsmath}	% Advanced maths commands
\usepackage{amssymb}	% Extra maths symbols

\newcommand{\artdeco}{\texttt{artDeco}}
\newcommand{\Artdeco}{\texttt{ArtDeco}}
\newcommand{\Planck}{\texttt{Planck}}

\newcommand{\lmax}{\ensuremath{\ell_{\text{max}}}}
\newcommand{\aslm}{$a_{s\ell m}$}
\newcommand{\atlm}{$a_{T\ell m}$}
\newcommand{\ve}[1]{\mathbf{#1}}

\providecommand{\sorthelp}[1]{}

\numberwithin{equation}{section} 

\title[Application of beam deconvolution to power spectrum estimation]
{Application of beam deconvolution technique to power spectrum estimation for CMB measurements}

\author[Keih\"anen et al.]
{E.~Keih\"anen$^{1,2}$\thanks{E-mail: elina.keihanen@helsinki.fi}
K.~Kiiveri$^{1,2}$
H.~Kurki-Suonio$^{1,2}$
\and
 M.~Reinecke$^{3}$
\\
$^{1}$ Department of Physics, Gustaf H\"allstr\"omin katu 2, 00014 University of Helsinki, Finland \\
$^{2}$ Helsinki Institute of Physics, Gustaf H\"allstr\"omin katu 2, 00014 University of Helsinki, Finland \\
$^{3}$ Max-Planck-Institut f\"ur Astrophysik, Karl-Schwarzschild-Str.~1, 85741 Garching, Germany
}

\date{Accepted XXX. Received YYY; in original form ZZZ}

\pubyear{2016}

\begin{document}

\label{firstpage}
\pagerange{\pageref{firstpage}--\pageref{lastpage}}
\maketitle

\begin{abstract}
We present two novel methods for the estimation of the angular power spectrum of cosmic microwave background
 (CMB) anisotropies. We assume an absolute CMB experiment with arbitrary asymmetric beams
 and arbitrary sky coverage.  The methods differ from earlier ones in that the power spectrum is estimated directly from time-ordered data,
without first compressing the data into a sky map, and they take into account the effect of asymmetric beams. 
In particular, they correct the beam-induced leakage from temperature to polarization.
The methods are applicable to a case where part of the sky has been masked out to remove foreground contamination,
 leaving a pure CMB signal, but incomplete sky coverage.
The first method (DQML) is derived as the optimal quadratic estimator, which simultaneously yields an unbiased 
spectrum estimate and minimizes its variance. We successfully apply it to multipoles up to $\ell$=200.
The second method is derived as a weak-signal approximation from the first one.
It yields an unbiased estimate for the full multipole range, but relaxes the requirement of 
minimal variance. We validate the methods with simulations for the 70 GHz channel of {\tt Planck} surveyor,
and demonstrate that we are able to correct the beam effects in the  $TT$, $EE$, $BB$, and $TE$ spectra 
up to multipole $\ell$=1500.   Together the two methods cover the complete multipole range with no gap in between.
\end{abstract}


%%%%%%%%%%%%%%%%%%%%%%%%%%%%%%%%%%%%%%%%%%%%%%%%%%

%%%%%%%%%%%%%%%%% BODY OF PAPER %%%%%%%%%%%%%%%%%%

%%%%%%%%%  SECTION %%%%%%%%%%
\section{Introduction}

Present-day cosmic microwave background (CMB) experiments require accurate methods 
for the estimation of the angular power spectrum of the CMB.
Several methods have been developed for this purpose.
The methods fall roughly into two categories,
according to the multipole range that they are applicable to.
The quadratic maximum likelihood (QML) estimator \citep{tegmark1997,tegmark2001}
gives an unbiased minimum-variance estimate for the low multipole range,
when given a CMB sky map as input.
A practical implementation of the method is the {\tt BolPol} estimator \citep{gruppuso2009},
which has been applied to multipoles up to $\ell$=32.

The high multipole regime requires different techniques.
To mention the ones most relevant for the current work,
the {\tt Master} \citep{hivon2002} method provides an unbiased estimate 
of the CMB temperature spectrum.
The method consists of computing the pseudo-$\cal C_\ell$ spectrum of the input map
through harmonic transform, and correcting it through a kernel matrix
that depends on sky coverage.
Similar methods have been developed for polarization 
\citep{kogut2003,chon2004,grain2009}.
For a more extensive review of power spectrum estimation (PSE)
methods we refer to \cite{gruppuso2009}.

Beam effects are typically not considered part of the power spectrum estimation problem,
but are taken into account at a later stage in the form of a beam window function.
A scalar beam window (for instance \citealt{mitra2010}) captures the average 
smoothing effect of the beam.
A scalar window function does not correct the leakage from temperature to
polarization.  Methods for correcting the latter include
matrix window function formalism \citep{planck2014-a05} and {\tt QuickPol} \citep{hivon2016}.

On the other hand, methods have been developed for full beam deconvolution at map level
\citep
{armitage-wandelt-2004,armitage-caplan2009,harrison-etal-2011,keihanen2012}.
Deconvolution map-making yields a sky map with a symmetrized effective beam.
This is achieved at the cost of more complicated noise structure,
which makes it more difficult to use the deconvolved maps for power spectrum estimation.
The impact of deconvolution on noise properties
has been discussed in \cite{keihanen2015}.  A low-multipole noise covariance matrix 
for deconvolved maps has been derived in the same paper.

In this paper we aim at combining the best of the two worlds. 
We present a method which combines full beam deconvolution 
with techniques of  power spectrum estimation.
The deconvolution technique is based on the \artdeco\ deconvolver \citep{keihanen2012}.
Our goal is to estimate the power spectrum directly from the time-ordered data (TOI),
without constructing  a sky map along the way,
at the same time eliminating the smoothing by an asymmetric beam.
In particular, we aim at removing the beam-induced leakage from 
temperature to polarization.

The situation we have in mind is the following.
Assume we have successfully performed component separation for our data.
We thus have an estimate of how much foreground emission there is in each sky pixel.
This is the starting point in pixel-based PSE methods as well.
Instead of proceeding with the foreground-cleaned CMB map, we go back to the TOI.
We can scan the foreground map into a TOI, convolving it with the known beam,
and subtract it from the original TOI. 
It is unlikely that the cleaning can be done with sufficient accuracy for all of the sky.
Therefore, the regions with strongest foreground emission are masked out.
We are thus left with a pure CMB time stream, but with incomplete sky coverage.

We aim at deriving the optimal quadratic method
which yields an unbiased estimate of the CMB spectrum, at the same time
minimizing its variance, when given a TOI stream as input.
The technique is inspired by the work of \cite{tegmark1997},
with the distinction that we include beam effects,
and, instead of maps, operate at TOI level. 
The required inputs beside the TOI itself are pointing information and 
known beam shapes.

We derive yet another method that is applicable to the full multipole range of interest.
Again the method yields an unbiased estimate of the CMB spectrum,
however not necessarily with minimal variance.
Formally we derive the high-ell method from the optimal method as a weak-signal approximation.
A key element in both methods is a kernel matrix that is applied to a raw spectrum
to correct simultaneously for beam effects and for incomplete sky coverage.

We validate both new methods with simulations.
As the basic simulation case we take the four-year data set of the \Planck\ LFI 
instrument's 70 GHz channel.

This paper is organized as follows.
We start by presenting the fiducial simulation case in Sect. 2.
We review the deconvolution map-making method of \artdeco\ in Sect. 3.
To set a reference point, we compare some pixel-based power spectrum estimation methods in Sect. 4.
The optimal PSE method is presented in Sect. \ref{sec:optimal},
and the high-multipole version in Sect. 6.
We validate the methods with simulations in Sect. 7.
In Sect. 8. we present an idea for future development, statistical estimation of 
the optimal kernel matrix.  Finally, we give our conclusions in Sect. 9.

%%%%%%%%%  SECTION %%%%%%%%%%

\section{Test case}
\label{sec:simulations}

We selected as test case the \Planck\ LFI instrument,
and its 2015 data release \citep{planck2014-a01}.
We run simulations on all LFI channels: 30, 44, and 70 GHz.
We consider both temperature and polarization.
From the cosmology point of view, the 70 GHz channel is the most interesting,
and we select it as the fiducial test case.
This channel offers a wide multipole range ($\ell$=0--1500)
and detectors beams asymmetric enough to demonstrate various beam effects.
It is also the computationally most demanding one, due to the size of the data set
and the wide multipole range covered.
The FWHM beam widths for the 70 GHz channel vary in the range 12.8-13.5',
allowing to estimate the power spectrum up to \lmax=1500.
The results shown in this paper are mainly based on the fiducial 70 GHz data set.
In some cases, where an effect under study is more prominent at another channel,
or the computation is very heavy, we show results from 30 GHz or 44 GHz.

We used the LevelS software package \citep{reinecke2006} to generate detector pointings,
to convolve the input sky with realistic beams, and to produce time-ordered data from it.
The simulations cover four years of mission, during which the sky is scanned eight times.
Because we are focusing on CMB power spectrum estimation,
we allowed some simplifications with respect to full \Planck\ data analysis.
We did not include systematics such as calibration errors,
and added white noise only. 
 The beams, however, were fully realistic, and included the main 
and intermediate sidelobe components of real \Planck\ beams \citep{planck2014-a05}.

The input sky contained the polarized CMB signal, but no foregrounds.
We applied the \Planck\ calibration masks \citep{planck2014-a06} to remove the galactic region 
from the data set.
This mimics a situation where an estimate of the foreground signal has been subtracted
from the TOI before passing it to power spectrum estimation,
and the regions of the strongest contamination have been cut out.
We are left with a data set that consists of pure CMB, but with incomplete sky coverage.
The sky coverage at 70 GHz is 89.67\%.
As input CMB sky we used one realization from the FFP8 simulation set \citep{planck2014-a14}

We also performed a series of pure white noise simulations,
which we processed through
the same deconvolution procedure as the CMB simulations.
The noise parameters were taken from \cite{planck2014-a02}.
Here we are assuming that the data has been destriped or otherwise cleaned of correlated
noise before the PSE phase, in sufficient degree that the residual noise can be considered
white. We discuss the effect of non-white residuals briefly in the results section.

The noise simulations serve several purposes. We use them to evaluate the noise bias,
which is subtracted from the estimated spectrum to reveal the actual CMB spectrum,
and to study the properties of residual noise. 
We utilize them also in the statistical kernel evaluation method presented 
in Sect. \ref{sec:statistical}.

%%%%%%%%%  SECTION %%%%%%%%%%

\section{Deconvolution map-making}
\label{sec:deconvolution}

\subsection{ArtDeco deconvolver}

Our work builds on the beam deconvolution formalism presented in
\cite{keihanen2012}. 
Consider a time-ordered data stream of the form
\begin{equation}
t_{i} = \sum_{s\ell m} A_{i,s\ell m}a_{s\ell m}+n_{i} . \label{toimodel}
\end{equation}
Here index $i$ labels the samples in the data stream,
$n_{i}$ is noise, and harmonic coefficients $a_{slm}$ represent the sky signal.
The spin index $s$ takes values $0$ and $\pm2$, 0 representing temperature
and $\pm2$ polarization.
There is a linear dependence between the sky $a_{s\ell m}$ and the data $t$,
collected in matrix $A$.
The explicit formula for $A$ is given in \cite{keihanen2012} as
\begin{equation}
  A_{i,s\ell m} = \sum_{k} b^{\ast}_{s\ell k}D^{\ell\ast}_{mk}(\omega_{i}) . \label{deconvmodel}
\end{equation}
Here $b_{s\ell k}$ is the harmonic expansion of the detector beam,  $D^\ell_{mk}$ 
is a Wigner matrix,  and $\omega_{i}$
 is the combination of three pointing angles $\{\theta,\phi,\psi\}$ that define the pointing and orientation 
 of the beam for sample $i$.

In deconvolution mapmaking we perform a linear fit to the TOI to obtain 
an estimate for the coefficients \aslm. From those one can construct a sky map,
where the effective beam is symmetric and Gaussian. 
The general deconvolution solution can be written in vector notation as
\begin{equation}
  \hat  {\ve a} = (A^\dagger N^{-1}A+S^{-1})^{-1} A^\dagger N^{-1}\ve t  . \label{deconvolution}
\end{equation}
Here $S$ represents an optional prior in the form of a first-guess CMB power spectrum.
Formally it is a diagonal matrix, with elements of the angular power spectrum spread along the diagonal.
In usual deconvolution mapmaking it is omitted,
but it has an important role in our power spectrum estimation method.

\subsection{Direct spectrum estimation}

The harmonic coefficients obtained from Eq. (\ref{deconvolution})
are not directly useful for the estimation of the CMB power spectrum,
if sky coverage is not complete.
The deconvolution as implemented in \artdeco\ can be performed under incomplete sky coverage as well,
and it works correctly in the sense that it recovers the sky map outside the mask.
However, the output \aslm\ do not represent the correct CMB spectrum.
One can imagine filling the missing sky region by an arbitrary CMB realization.
Each possible realization corresponds to a different set of \aslm, each of which gives 
the correct sky map in the region outside the mask. 
From these realizations, \artdeco\ arbitrarily picks one, depending on the starting point
of the conjugate-gradient iteration.
Without additional information it is not possible to identify the correct realization.

\begin{figure}
\includegraphics[width=8.8cm]{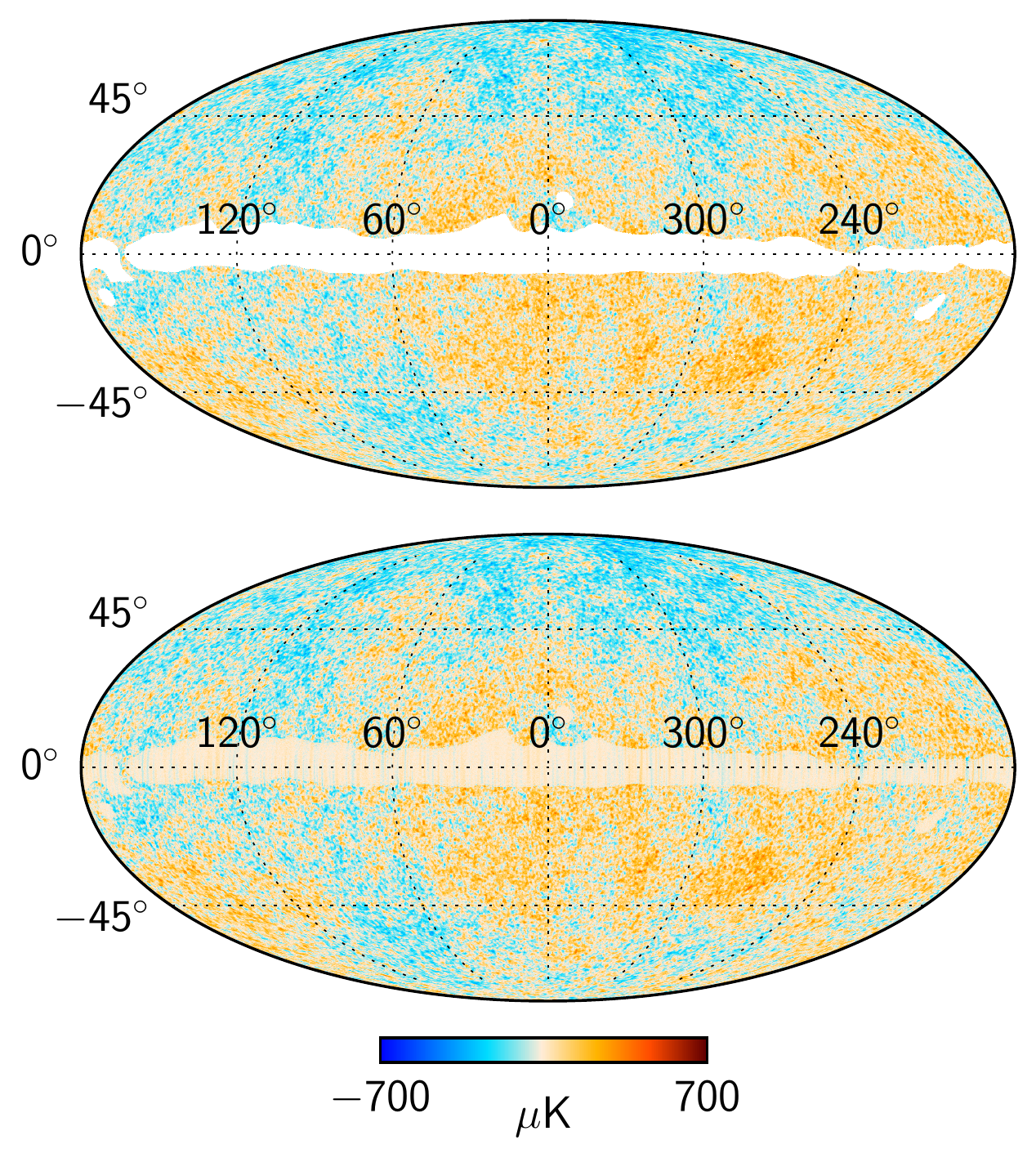}
\caption{
Simulated CMB sky, seen through \Planck\ LFI 70 GHz beams.  
Top: Map binned directly from the time-ordered data.  
Galaxy and strong point sources are masked out,
leaving a pure CMB signal.
Bottom: Beam-deconvolved version of  the same map,
constructed through harmonic expansion
from deconvolved \aslm\ coefficients.
Deconvolution restores the CMB signal in the observed region,
but creates artefacts in the unobserved region.}
\label{fig:binmap}
\end{figure}

We demonstrate deconvolution map-making in Fig. \ref{fig:binmap}.
We show two versions of the temperature map,
constructed from a simulated timeline with \Planck\ 70 GHz beams.
We show first the binned map, which is constructed by
coadding the TOI samples into a two-dimensional sky,
without attempt to correct for beam shapes.
Below it we show the deconvolved version of the same map.
To construct the latter we run the data through \artdeco\ deconvolver
to obtain a beam-free \aslm\ expansion of the sky,
 reconstruct the sky map through harmonic transform, and apply Gaussian smoothing
with FWHM=12'. 
Both maps make use of Healpix pixelization at resolution $N_{\rm side}=1024$ (3,5').
Despite there being no data available from the masked region,
deconvolution inserts signal there as well.

\begin{figure}
\includegraphics[width=8.8cm]{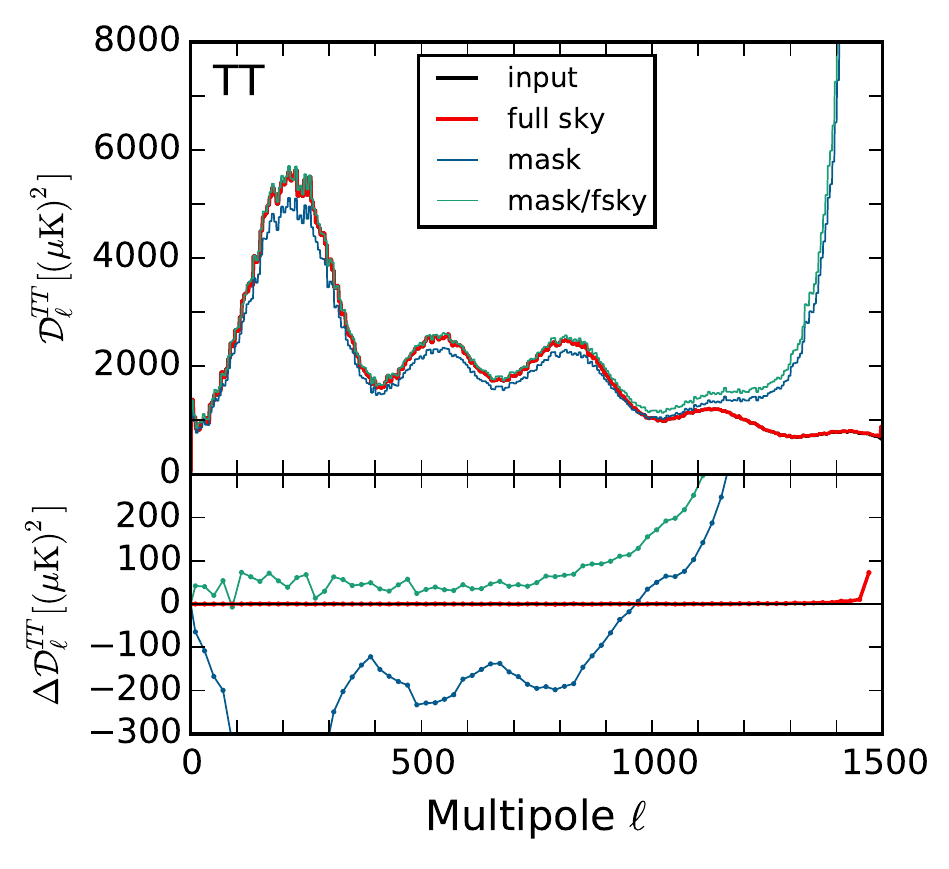}
\caption{
$TT$ spectrum as constructed directly from deconvolved \atlm\ coefficients.
Simulation data consists of pure CMB signal, convolved by \Planck\ 70 GHz beams.
Here, as throughout the paper,
we plot the quantity ${\cal D}_\ell={\cal C}_\ell\ell(\ell+1)/(2\pi)$,
which shows the high-multipole structure more clearly.
In the lower panel we plot the absolute error in the same units.
In the ideal case with 100\% sky coverage, 
the deconvolved \aslm\ coefficients directly give a reliable estimate of the CMB spectrum.
This is no longer true when deconvolution is applied to a data set where the 
part of the sky is masked out. Scaling the spectrum by the inverse of sky fraction
 $f_\mathrm{sky}=0.8967$
provides a rough correction (green), but still leaves a systematic error of the order of 1\%.  
 }
\label{fig:fullsky}
\end{figure}

In Fig. \ref{fig:fullsky} we show the raw TT spectrum constructed directly from the deconvolved
harmonic coefficients as
\begin{equation}
\hat {\cal C}^{TT}_\ell=\frac{1}{2\ell+1}\sum_m a_{Tlm}^\ast a_{Tlm}
\end{equation}
for full sky, and in presence of a mask.
In the lower panel we show the absolute error. 
Throughout the paper we follow the convention where we compare the recovered spectrum
to the spectrum of the single CMB realization that was used as input to the simulation.
We plot the quantity ${\cal D}_\ell={\cal C}_\ell\ell(\ell+1)/(2\pi)$, which shows the high multipole 
range more clearly.
As measure of error we plot the absolute error, calculated as the difference between 
the recovered spectrum and the input spectrum.  This allows use to use the same error measure 
for the $TE$ cross-spectrum, for which the relative error is an inconvenient measure.

We see in Fig. \ref{fig:fullsky} that in the ideal case where the sky coverage is complete,
and no noise is included in the simulations,
deconvolution recovers the true CMB spectrum almost perfectly.
The situation changes when a mask is applied. Direct deconvolution yields a spectrum
that falls below the input at low multipoles, but rises steeply at high multipoles.
The effect at low multipoles corresponds roughly to scaling the spectrum by the sky coverage,
and can as first approximation be corrected for by inverse scaling.
There remains, however, a 1\% error, which is unacceptably high for present-day precision cosmology.
More accurate methods are thus needed.
At high multipoles (above $\ell$=800) the deconvolved spectrum becomes useless.

Even if we cannot determine the individual harmonic coefficients,
we still may be able to estimate their spectrum.
The main goal of this paper is to derive a method that extracts the spectral information
in an optimal way, given the pre-known beam shapes and known noise level.

%%%%%%  FIGURES %%%%%%%%%%%%%%

%%%%%%%%%% SECTION  %%%%%%%%%%%%%%%%%%%%%%

\section{Beam correction at the power spectrum level}
\label{sec:pixelmethods}

To put the new methods into context, we first compare some available map-based methods 
for power spectrum estimation.
We use again the simulated \Planck\ 70 GHz data set, with realistic beams. 
A Galactic mask with $f_0=0.8967$ sky coverage is applied.
The methods discussed here take as starting point the (I,Q,U)
Stokes map triplet binned from time-ordered data.  
The operation of constructing the map from the TOI
utilizes the known direction of polarization sensitivity for each detector,
encoded in parameter $\psi_{\rm pol}$, but no other information on beam properties.

\begin{figure*}
\includegraphics[width=18cm]{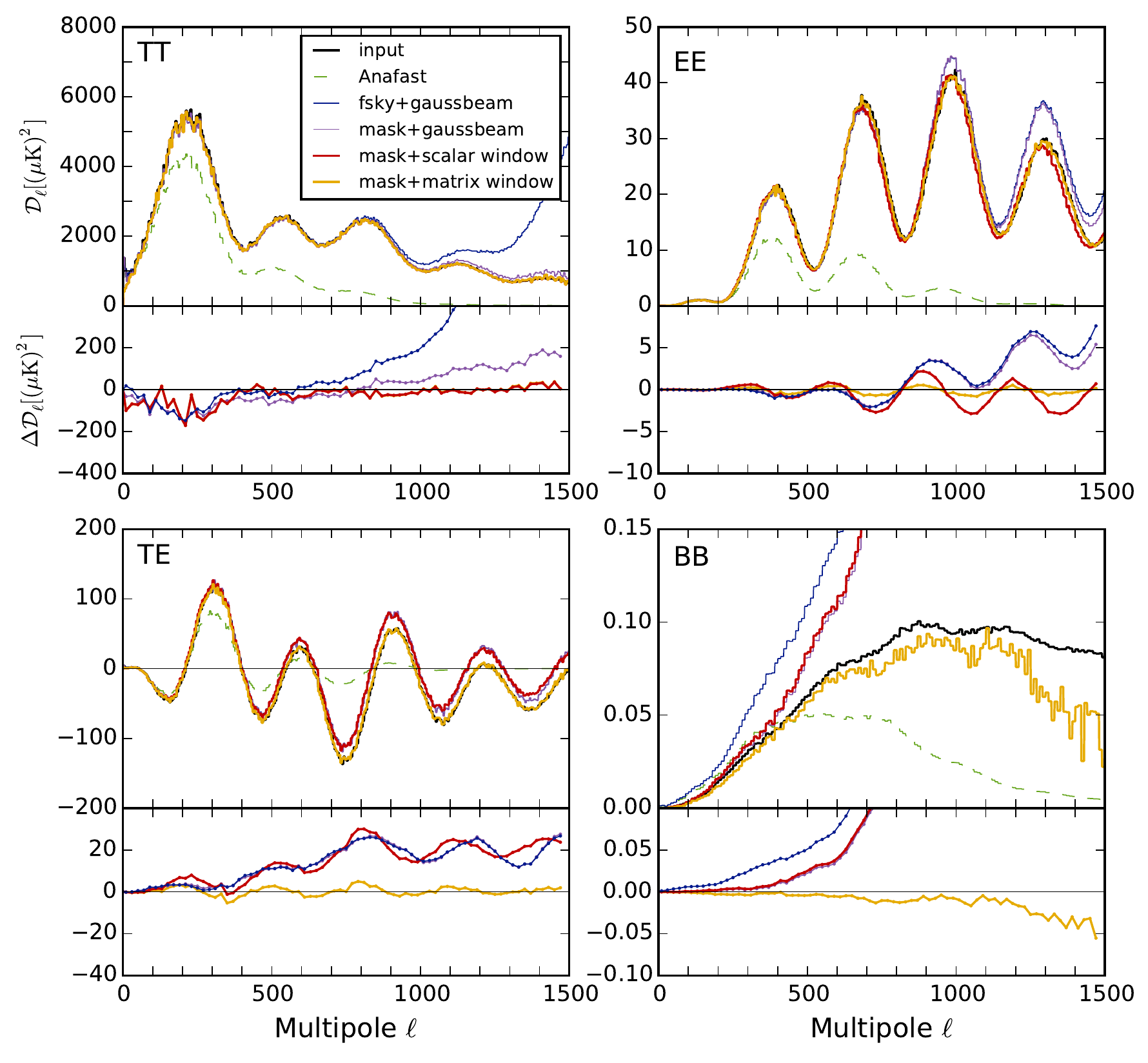}
\caption{
Comparison of selected pixel-based methods for power spectrum estimation,
applied to the fiducial 70 GHz simulation.
We show the spectrum of the input CMB realization,
the un-corrected Anafast spectrum (dashed),
and a series of estimates with increasing complexity:
Scaling by sky fraction $f_\mathrm{sky}$ and Gaussian beam window,
and {\tt PolSpice} mask correction combined with Gaussian beam, scalar beam window function
or matrix beam window function.
The difference between scalar and matrix window functions reveals the level of beam-induced leakage
from temperature to polarization.
The spectra are averaged over 5 adjacent multipoles to reduce scatter,
except for $BB$ which is averaged over 10 multipoles.
The absolute error given in the lower panel is averaged over 20 multipoles.
}
\label{fig:corrections}
\end{figure*}

The pseudo spectrum constructed from the map is suppressed both by the beam shape
and due to the incomplete sky coverage. We refer to this initial spectrum as Anafast
spectrum, according to the standard {\tt HEALPix} \citep{gorski2005} tool. 
We show in Fig. \ref{fig:corrections} the Anafast spectrum along with
various corrections.   
The simplest correction consists of dividing the Anafast spectrum by the sky fraction $f_\mathrm{sky}$,
and by a Gaussian beam window estimate. The FWHM width of the Gaussian window was
taken from \cite{planck2014-a05} to be 13.315'.
This naive correction works reasonably well at low multipoles, but overestimates all the spectra towards
higher multipoles.

%%%%%%%%%%%%%%%%%%%%%%%%%%%%%%%%%%%%

%%%%%%%%%%%%%%%%%%%%%%%%%%%%%%%%%%%%

Several more sophisticated methods exist, that take into account the exact mask shape and correct 
the effect of the incomplete sky coverage.
We use the publicly available {\tt PolSpice}\footnote{\tt www2.iap.fr/users/hivon/software/PolSpice/} 
code \citep{chon2004} to apply the correction to our data set.
we then apply the Gaussian beam window to correct for the beam smoothing.
The mask correction improves the spectrum estimate, especially in the $BB$ spectrum, 
and in $TT$ at high multipoles.

We may further improve the spectrum estimate by replacing the Gaussian beam approximation 
by a proper beam window computed from full beam information.
The scalar beam window function $W^u_\ell$ can be obtained from MC simulations using 
the known instrument beam as
\begin{equation}
	W^u_\ell \equiv \frac{ \langle \tilde {\cal C}^u_\ell \rangle_\mathrm{MC} }{ {\cal C}^u_\ell } \,,
\label{eq:scalar_W}
\end{equation}	
where ${\cal C}^u_\ell$ ($u=TT, EE\ldots$) is some input spectrum used for the simulations, 
$\tilde {\cal C}^u_\ell$ 
is the output (Anafast) 
full-sky spectrum, and $\langle \rangle_\mathrm{MC} $ is an average over many realizations.  
This method is not suitable for treating leakage between temperature and polarization signals, 
so to treat just the smoothing
effect we use Eq.~(\ref{eq:scalar_W}) for $u = TT, EE, BB$ with separate $T$-only, $E$-only, and $B$-only simulations. 
For $u = TE, TB, EB$ we then use
\begin{equation}
	W^{TE}_\ell = \sqrt{W^{TT}_\ell W^{EE}_\ell} \quad\mbox{etc.}
\end{equation}
The angular power spectrum estimate $\hat {\cal C}^u_\ell$ for spectral component $u$ is obtained 
from the Anafast spectrum  
by dividing by the corresponding beam window.
	
For correcting the beam-induced leakage between temperature and polarization signals, a matrix window function method
was introduced in \cite{planck2014-a05}.  Here the estimate $\hat {\cal C}_\ell$ is obtained from the spectrum 
$\tilde {\cal C}_\ell$ of the sky map as
\begin{equation}
	\hat {\cal C}_\ell = W^{-1}_\ell \tilde {\cal C}_\ell \,,
\end{equation}
where the ${\cal C}_\ell$ are 6-component vectors with components ${\cal C}^u_\ell$ and the $W_\ell$ is 
a $6\times6$ matrix with components $W^{uu'}_\ell$.  The components $W^{uu'}_\ell$ of the matrix window function 
are obtained from 
$T$-only, $E$-only, and $B$-only MC simulations as described in \citep{planck2014-a05}.  
For correcting the effect of the incomplete sky coverage prior to the beam correction, we again use the {\tt PolSpice} method.

The difference between the results obtained with scalar beam window and with matrix beam window 
is an indicator of the level of  leakage between spectral components,
an effect that cannot be corrected for through a scalar window function.
As can be expected, the difference is significant in polarization components,
but negligible in temperature.

Of all the correction methods compared here, the last one ({\tt PolSpice} mask correction combined
with matrix beam window) provides the best correction.
This sets a reference level for the accuracy that we must require from our deconvolution-based
spectrum estimation method.

%%%%%%%%%%%% SECTION  %%%%%%%%%%%%%%%%%%%%

\section{Optimal power spectrum estimation}
\label{sec:optimal}

\subsection{Notation}

We now set out to derive a power spectrum estimation method,
which corrects simultaneously the effect of the incomplete sky coverage,
and that of beam smoothing, including beam-induced leakage effects. 
We start by introducing some formalism used throughout this work.

Consider first the properties of the CMB \aslm\ elements.
We denote by $S$ their covariance,
\begin{equation}
  \langle a_{s\ell m}a_{s'\ell'm'}^{\ast}\rangle = S_{s\ell m,s'\ell'm'}.
\end{equation}
All elements except the ones with $\ell=\ell'$ and $m=m'$ vanish.
The non-zero elements are linear combinations of elements of the CMB power spectra
${\cal C}_{TT}$, ${\cal C}_{EE}$, ${\cal C}_{BB}$, ${\cal C}_{TE}$, ${\cal C}_{TB}$, and ${\cal C}_{EB}$.
Parity symmetry indicates ${\cal C}_{TB}={\cal C}_{EB}=0$.
To keep the formalism as general as possible, we retain these components as well.
The unwanted elements can be dropped at any later stage.
 
The spin harmonics are related to the $T,E,B$ harmonic components through
\begin{eqnarray}
a_{0\ell m} &=& a_{T\ell m} \nonumber \\
a_{2\ell m} &=& -(a_{E\ell m}+ia_{B\ell m}) \\
a_{-2\ell m} &=& -(a_{E\ell m}-ia_{B\ell m})  . \nonumber 
\end{eqnarray}
From this we can derive the linear relation between the spectral components,
\begin{eqnarray}
 \langle a_{0lm}a_{0l'm'}^{\ast} \rangle 
 &=& {\cal C}^{TT}_\ell  \delta_{\ell\ell'}\delta_{mm'} \label{spinspectra}  \\
  \langle a_{2\ell m}a_{2\ell'm'}^{\ast} \rangle 
 &=&  ({\cal C}^{EE}_\ell +{\cal C}^{BB}_\ell) \delta_{\ell\ell'}\delta_{mm'}  \nonumber \\
 \langle a_{-2\ell m}a_{-2\ell'm'}^{\ast} \rangle 
 &=&  ({\cal C}^{EE}_\ell +{\cal C}^{BB}_\ell) \delta_{\ell\ell'}\delta_{mm'}  \nonumber \\
  \langle a_{2\ell m}a_{-2\ell'm'}^{\ast} \rangle 
  &=& ({\cal C}^{EE}_\ell -{\cal C}^{BB}_\ell +i2{\cal C}^{EB}_\ell) \delta_{\ell\ell'}\delta_{mm'}  \nonumber \\
   \langle a_{0\ell m}a_{2\ell'm'}^{\ast} \rangle
 &=& -({\cal C}^{TE}_\ell-i{\cal C}^{TB}_\ell) \delta_{\ell\ell'}\delta_{mm'}            \nonumber \\
  \langle a_{0lm}a_{-2l'm'}^{\ast} \rangle
 &=& -({\cal C}^{TE}_\ell+i{\cal C}^{TB}_\ell) \delta_{\ell\ell'}\delta_{mm'}    .        \nonumber
\end{eqnarray}
We can write this as
\begin{equation}
   \langle a_{s\ell m}a_{s'\ell'm'}^{\ast} \rangle 
    = \delta_{\ell\ell'}\delta_{mm'} \sum_u g^{u}_{ss'} {\cal C}_{ul} , \label{gdef}
\end{equation}
where $u$ labels the spectra ($u$=$TT$, $EE$, $BB$, $TE$, $TB$, $EB$), and
$g$ is a $3\times3\times6$ object whose non-zero elements are
\begin{eqnarray}
   g_{00}^{TT} &=& 1  \nonumber\\
   g_{0,-2}^{TE} = g_{0,2}^{TE} = g_{-2,0}^{TE} = g_{2,0}^{TE} &=& -1 \nonumber\\
   g_{2,2}^{EE} = g_{2,-2}^{EE} = g_{-2,2}^{EE} = g_{-2,-2}^{EE} &=& 1 \label{gmatrix}\\
   g_{2,2}^{BB} = g_{-2,-2}^{BB} = -g_{2,-2}^{BB} = -g_{-2,2}^{BB} &=& 1   \nonumber \\
   g_{0,2}^{TB}  = g_{-2,0}^{TB} = -g_{0,-2}^{TB} =  -g_{2,0}^{TB} &=& i \nonumber \\
   g_{2,-2}^{EB} = -g_{-2,2}^{EB} &=& 2i . \nonumber 
\end{eqnarray}
We further define object $G$ as
\begin{equation}
   G^{u\ell''}_{s\ell m,s'\ell'm'} = \delta_{\ell\ell''}\delta_{\ell\ell'}\delta_{mm'}g^{u}_{ss'} . \label{shat}
\end{equation}
With this definition the relations of Eq. (\ref{spinspectra}) can be written in the compact form
\begin{equation}
  \langle a_{s\ell m}a_{s'\ell'm'}^{\ast} \rangle 
  = \sum_{u\ell''} {\cal C}_{u\ell''} G^{u\ell''}_{s\ell m,s'\ell'm'}.  \label{smatrix}
\end{equation}

We introduce combined indices that allow a still more concise notation.  
Index $n$ represents the triplet of indices $n=\{s,\ell,m\}$
and index $L$ the combination of indices $L=\{u,\ell\}$.
The time-ordered data of Eq. (\ref{toimodel})  becomes
\begin{equation}
t_{i} = \sum_{n} A_{in}a_{n}+n_{i}  , \label{toimodel2}
\end{equation}
and Eq. (\ref{smatrix}) becomes
\begin{equation}
   S_{nn'} =  \langle a_{n}a_{n'}^\ast \rangle = \sum_{L} {\cal C}_{L} G^{L}_{nn'}.  \label{smatrix2}
\end{equation}

It is useful to note that the standard operation of 
calculating the spectrum of a given harmonic vector $a_{slm}$
is formally given as
\begin{equation}
\hat {\cal C}_L = \left[\sum_{nn'}G^{\ast L}_{nn'}G^{L'}_{nn'}\right]^{-1} 
\sum_{nn'}a_n G^L_{nn'} a^\ast_{n'} .  \label{almspec}
\end{equation}
When written out, this gives, for instance, for the $TT$ spectrum the usual formula
\begin{equation}
\hat {\cal C}^{TT}_\ell = \frac{1}{2\ell+1} \sum_m |a_{0\ell m}|^2.
\end{equation}
Matrix $G$ represents a quadratic sum of two \aslm\ vectors, 
and the factor in the brackets takes care of normalization.

\subsection{Finding the optimal quadratic method}

We aim at finding a quadratic method, represented by matrix $E^{L}$, 
which gives an optimal estimate 
for the power spectrum  ${\cal C}_{L}$, given known beams and time-ordered data.
The data is assumed to consist of CMB signal and noise.
We follow the example of the Quadratic Maximum Likelihood (QML) 
analysis of \cite{tegmark1997},
the major difference being that we are dealing with full time streams instead of sky maps.
We refer to the new method as DQML (Deconvolution Quadratic Maximum Likelihood).

The optimal estimate must fulfil two requirements:  \\
(i) The solution is unbiased (as ensemble average). \\ 
(ii) Under condition (i) the solution minimizes the variance.

Consider first condition (i).
Given a time stream $t$,
we want to find a (symmetric) matrix $E^{L}_{ij}$, and constant $\beta_{L}$, such that
\begin{equation}
  \hat {\cal C}_{L} = \sum_{ij} t_{i}E^{L}_{ij}t_{j} -\beta_{L}\label{edef}
\end{equation}
gives an unbiased estimate of the true power spectrum ${\cal C}^{L}$,
that is
\begin{equation}
  \langle \hat {\cal C}_{L}\rangle = {\cal C}_{L}.  \label{unbiaseddef}
\end{equation}
The brackets denote an ensemble average over noise realizations and 
realizations of $a_{n}$ for a given power spectrum.

Inserting (\ref{toimodel2}) into (\ref{edef}) and averaging we obtain,
assuming that noise and signal are independent,
\begin{eqnarray}
  \langle \hat {\cal C}_{L} \rangle &=& \sum_{ij}E^{L}_{ij}
  \left( \sum_{nn'}A_{jn}\langle a_{n}a_{n'}^{\ast}  \rangle  A^{\ast}_{in'}
      +\langle n_{j} n^{}_{i}\rangle  \right) -\beta_{L}  \label{derunbiased} \\
  &=& \sum_{ij}E^{L}_{ij} \left( \sum_{nn'}A_{jn} A^{\ast}_{in'} \sum_{L'} {\cal C}_{L'}G^{L'}_{nn'}
      +N_{ji}\right) -\beta_{L} . \nonumber
\end{eqnarray}
We have denoted the noise covariance in TOI domain by
\begin{equation}
  \langle n_{i} n_{j} \rangle = N_{ij} . \end{equation}
If noise is white, this is a simple diagonal matrix.

The requirement of Eq. (\ref{unbiaseddef}) now splits into two conditions.
From the noise part we get a relation for the noise bias $\beta$,
\begin{equation}
 \beta_{L} = \sum_{ij} E^{L}_{ij}N_{ji} ,
\end{equation}
which we can evaluate once we have found out  $E$.
The signal part gives the condition
\begin{equation}
  \sum_{ij}E^{L}_{ij} \sum_{nn'}A_{jn} A^{\ast}_{in'} G^{L'}_{nn'}  
 \delta_{LL'}. \label{unbiased}
\end{equation}
These can be written in matrix formalism as
\begin{equation}
 \beta_{L} = \mathrm {Tr}(E^{L}N),
\end{equation}
and
\begin{equation}
  \mathrm {Tr}(E^{L}AG^{L'}A^{\dagger}) = \delta_{LL'}. \label{unbiased_mat}
\end{equation}
When using matrix formalism we are interpreting $E$ and $N$ as matrices
with $i,j$ as row and column index.
Each value of index $L$ thus identifies one matrix. The trace operation
applies to indices $i,j$. Similarly, $G$ is interpreted as a matrix with $n,n'$ 
as column and row index.

The condition of Eq. (\ref{unbiased_mat}) does not yet uniquely determine $E$.
Consider then condition (ii).
Among all $E^{L}$ that satisfy condition (\ref{unbiased_mat}), 
we want to find the one which minimizes the variance,
\begin{equation}
  V= \langle \hat{\cal C}_{L}^{2}\rangle -\langle \hat{\cal C}_{L}\rangle^{2}.
\end{equation}
We can write ($\beta$ cancels out)
\begin{eqnarray}
 V&=& \langle \sum_{ij}t_{i}E^{L}_{ij}t_{j}  \sum_{i'j'}t_{i'}E^{L}_{i'j'}t_{j'}\rangle
 - \langle\sum_{ij}t_{i}E^{L}_{ij}t_{j}\rangle
 \langle\sum_{i'j'}t_{i'}E^{L}_{i'j'}t_{j'}\rangle \nonumber \\
 &=& \sum_{ii'jj'}E^{L}_{ij}E^{L}_{i'j'}(\langle t_{i}t_{j}t_{i'}t_{j'}\rangle
      -\langle t_{i}t_{j}\rangle \langle t_{i'}t_{j'}\rangle) \nonumber  \\
 &=& \sum_{ii'jj'}E^{L}_{ij}E^{L}_{i'j'}
 ( \langle t_{i}t_{i'}\rangle \langle t_{j}t_{j'}\rangle
      +\langle t_{i}t_{j'}\rangle \langle t_{i'}t_{j}\rangle) \nonumber \\
 &=& \sum_{ii'jj'}E^{L}_{ij}E^{L}_{i'j'} (C_{ii'}C_{jj'}+C_{ij'}C_{i'j})  . \label{variance}
\end{eqnarray}
On the third line we have used the property of Gaussian random variates:
\begin{equation}
  \langle x_1x_2x_3x_4\rangle 
  = \langle x_1x_2\rangle\langle x_3x_4\rangle
  +\langle x_1x_3\rangle\langle x_2x_4\rangle
  +\langle x_1x_4\rangle\langle x_2x_3\rangle .
\end{equation}
Matrix $C$ on the last line represents the time-domain covariance,
including both CMB and noise,
and is given by
\begin{eqnarray}
  C_{ii'} = \langle t_{i}t_{i'}\rangle 
  &=& \sum_{nn'}A_{in}\langle a_{n}a_{n'}^\ast\rangle A^\ast_{i'n'}
    +\langle n_{i}n_{i'} \rangle   \nonumber \\
  &=& \sum_{nn'} A_{in}S_{nn'}A^{\ast}_{i'n'} +N_{ii'} ,
\end{eqnarray}
or, in matrix notation,
\begin{equation}
  C =  A S A^{\dagger} +N .  \label{cmat}
\end{equation}

We want to minimize the variance (\ref{variance}) under condition (\ref{unbiased_mat}).
We use the technique of Lagrange multipliers (see Tegmark's work).
The technique is intuitively understood as follows.
The constraint of Eq. (\ref{unbiased_mat}) picks a surface in the multidimensional space of $E_{ij}$.
At a local minimum of $V$ on this surface, the gradient of $V$ does not have a component 
along the surface. This is equivalent to saying that the gradient is included in the subspace spanned
by $AG^\dagger A^\dagger$.
The proportionality constants are the Lagrange multipliers, and are adjusted 
so as to fulfil the constraint (\ref{unbiased_mat}).

We take the gradient of  $V$ with respect to $E^{L}_{ij}$,
and write
\begin{equation}
\begin{split}
  2\sum_{i'j'}(C_{ii'}C_{jj'}+C_{ij'}C_{i'j})E^{L}_{i'j'} = \\
  4\sum_{L'}\lambda_{LL'}\sum_{nn'}A_{in}G^{L'\dagger}_{nn'}A^{\ast}_{jn'}.  \label{lagrangefirst}
 \end{split}
\end{equation}
We have arbitrarily scaled the right-hand-side by 4 (we can think of it as being absorbed in $\lambda$).
Rearranging the indices, and remembering that $E$ and $C$ are symmetric, we
find that the two terms on the left hand side are identical. The factor of 4 cancels out,
and  Eq. (\ref{lagrangefirst}) becomes in matrix notation
\begin{equation}
  C E^{L}C = \sum_{L'}\lambda_{LL'}AG^{L'\dagger}A^{\dagger} .
  \end{equation}
Since $C$ is a covariance matrix, it is invertible.
We can thus solve for $E$ as
\begin{equation}
  E^{L} = \sum_{L'}\lambda_{LL'} C^{-1}AG^{L'\dagger}A^{\dagger}C^{-1} .
\end{equation}
This gives the optimal quadratic method we are looking for.
We yet have to determine the Lagrange coefficients $\lambda$.
Inserting $E$ to Eq. (\ref{unbiased_mat}) we obtain
\begin{equation}
  \sum_{L'}\lambda_{LL'} \mathrm{Tr}\left[
  A^{\dagger}C^{-1}AG^{L'\dagger}
  A^{\dagger}C^{-1} AG^{L''} \right] = \delta_{LL''} .
\end{equation}
From this we can solve the coefficients $\lambda$,
by inverting the matrix defined by the trace formula.
The minimum-variance estimate for the power spectrum
is then obtained as
\begin{equation}
  \hat{\cal C}_L = \sum_{L'}\lambda_{LL'} t^TC^{-1}AG^{L'\dagger}A^{\dagger}C^{-1} t .
\end{equation}

We now have a formal solution for our power spectrum estimation problem.
The solution in its present form is still impractical,
since it involves inverting the the huge 
TOI covariance $C$. 
We need to work the solution yet into a more practical form.
With help of Sherman-Morrison formula and Eq. (\ref{cmat}) we can write
\begin{equation}
  A^\dagger C^{-1}t = S^{-1}(S^{-1}+A^{\dagger}N^{-1}A)^{-1} A^{\dagger}N^{-1}t. 
\end{equation}
Apart from factor $S^{-1}$ in front, this is the usual deconvolution equation of Eq. (\ref{deconvolution}).
With a similar technique we get 
\begin{equation}
  A^{\dagger}C^{-1}A =  S^{-1} - S^{-1}(S^{-1}+A^{\dagger}N^{-1}A)^{-1} S^{-1} .
  \label{mmatrix}
\end{equation}
We have replaced the TOI covariance by the much smaller deconvolution matrix, 
which has the rank equal  to the number of harmonic coefficients $3(\ell_\text{max}-1)^2$.

\subsection{DQML algorithm}
\label{sec:algorithm}

We are now ready to collect the results into a recipe for optimal power spectrum estimation.

\begin{enumerate}
\item  Run the usual beam deconvolution procedure on the TOI,
using some first estimate of the power spectrum as a prior,
\begin{equation}
  \hat {\ve a} = (S^{-1}+A^\dagger N^{-1}A)^{-1} A^\dagger N^{-1}\ve t. \label{adeconvolved}
\end{equation}
This step is performed by the \artdeco\ code.

\item Construct the raw spectrum as
\begin{equation}
  P_{L}  =  {\hat {\ve a}}^{\dagger} S^{-1} G^{L\dagger} S^{-1} \hat {\ve a} .  \label{Praw}
\end{equation}
Apart from normalization, matrix $G^L$ represents the familiar squared-sum operation of constructing 
the spectrum of a harmonic vector $\hat {\ve a}$.
Explicit from for $G$ is given by Eq. (\ref{shat}).
Matrix $S$ represents scaling by the CMB spectrum used as prior. Formally it is a diagonal matrix
with elements of the prior on the diagonal.

\item Construct the kernel matrix 
\begin{equation}
  \Lambda_{LL'} = \mathrm{Tr}\left[M G^{L}M^\dagger G^{\dagger L'}\right]  , \label{Lambda}
\end{equation}
where $M$ is given by
\begin{equation}
  M = S^{-1} - S^{-1}(S^{-1}+A^\dagger N^{-1}A)^{-1} S^{-1}  . \label{Mdef}
\end{equation}
This is the computationally heavy step. It involves constructing explicitly the full
deconvolution matrix. Routines for that exist in the \artdeco\ code.

\item Apply the kernel to the raw spectrum
\begin{equation}
\hat{\cal C}_L = \sum_{L'}\Lambda^{-1}_{LL'} P_{L'}.
\end{equation}

\item If noise is present, evaluate the noise bias $\beta_L$ 
through Monte Carlo simulations,
and subtract the bias from the estimate of step 4.
\end{enumerate}

\bigskip
The solution contains an estimaye of the CMB spectrum ${\cal C}_L$ inside the prior $S$.
It is not known, since it is exactly the quantity we are trying to determine.
It is important to note however, that $S$ only enters in the part of 
calculation where we minimized the variance.
If our estimate of $S$ is inaccurate, the solution will not be that of minimal variance 
but {the solution is still unbiased}. 

\subsection{Implementation aspects}

When implementing the method we made a couple of further rearrangements.
We note that the prior $S$ is independent of index $m$. From that, and from the special structure
of $G$, it follows that $S^{-1}$ can be taken out from Eq. (\ref{Praw}) and transferred inside the kernel of Eq. (\ref{Lambda}).
Further, we replace the operation $G$ in Eq. (\ref{Praw}) by the normalized version of Eq. (\ref{almspec}),
and apply the name normalization to the left-hand-side of the kernel.
The modified kernel is dimensionless, but no more symmetric. 
The raw spectrum of step 2 is replaced by
\begin{equation}
 \tilde P_{L}  =   \left[ \mathrm{Tr}( G^{L}G^{L'\dagger} )\right]^{-1} {\hat {\ve a}}^{\dagger} G^{L\dagger}\hat {\ve a}  . \label{Praw2}
\end{equation}
Equation (\ref{Praw2}) represents the standard operation of constructing the spectrum of a harmonic
vector, and is equivalent to Eq. (\ref{almspec}), only written in matrix notation.
The corresponding kernel is
\begin{equation}
 \tilde \Lambda_{LL'} =   \left[ \mathrm{Tr} G^{L}G^{L'\dagger} )\right]^{-1} 
  \mathrm{Tr}\left[\tilde M G^{L}\tilde M^\dagger G^{\dagger L'}\right] , \label{Lambda2}
\end{equation}
where $\tilde M$ is given by
\begin{equation}
  \tilde M = I -(S^{-1}+A^\dagger N^{-1}A)^{-1} S^{-1}  . \label{Mdef2}
\end{equation}

\begin{figure}
\includegraphics[width=8.8cm]{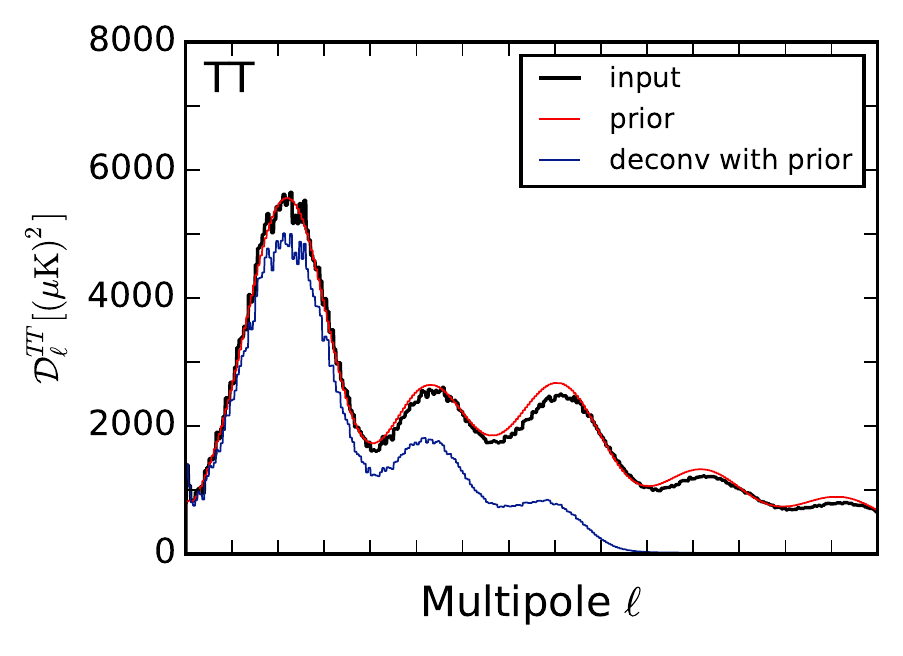}
\caption{
70 GHz $TT$ spectrum used as input in simulations and
the one used as prior. Deconvolution with prior yields a suppressed spectrum estimate.
The suppressed spectrum acts as input to the PSE estimation method of Sect. \ref{sec:optimal}.
}
\label{fig:prior}
\end{figure}

The modified algorithm is fully equivalent to the original one,
but is a little more intuitive. The right-hand-side ``raw'' spectrum has 
the physical interpretation as the spectrum of the deconvolved harmonic coefficients,
and can be constructed using standard tools.  For us remains the task of constructing the kernel $\tilde\Lambda$. 
The raw $TT$ spectrum in our simulations is depicted in Fig \ref{fig:prior}.
Because of the prior, it is suppressed with respect to the input spectrum.
A dimensionless kernel is then applied to the spectrum to correct for the suppression.

The construction of the kernel in step 3 is the computationally heaviest part in the algorithm.
To construct matrix $M$, we need to construct and invert the full deconvolution matrix.
\Artdeco\ contains routines that allow to evaluate any element of the matrix.
Storing and inverting the matrix, however, soon becomes unfeasible, as the matrix rank increases
as proportional to 3(\lmax+1)$^2$. We were able to compute the full kernel
up to \lmax=200.  The construction of the kernel took 4 hours on 1440 CPUs.

Constructing the raw spectrum is a lighter task.
We run the deconvolution at full resolution, then extract the multipoles
up to the \lmax\ of the kernel.

\subsection{Strong signal limit}
\label{sec:strongsignal}

It is instructive to look at the optimal solution in the ideal case,
where the data covers the full sky, and noise is absent or negligible.
As we saw in Sect. \ref{sec:deconvolution}, in this case direct
deconvolution recovers the input spectrum to a very high accuracy.
If the CMB signal is very strong compared with noise,
 $S^{-1}$ becomes negligible compared with $A^{\dagger}N^{-1}A$,
 and we can make the approximation 
\begin{equation}
  \tilde M \approx  I .
\end{equation}
The requirement of complete sky coverage is essential for 
matrix  $A^{\dagger}N^{-1}A$ to be invertible.
The kernel reduces into a unity matrix $\tilde\Lambda=I$,
and we see that the spectrum of the deconvolved
 $\hat a_{s\ell m}$ coefficients is indeed the optimal solution.

%%%%%%%%%%%%%%%%%%%%%%%%%%%%%%%%%%%%%%%%%%%%%

\section{High multipoles: weak signal approximation}
\label{sec:noisedom}

\subsection{Weak signal limit}

The optimal method of Sect. \ref{sec:optimal} is only feasible at a limited multipole range.
To cover the whole multipole range of interest (in our fiducial simulation up to \lmax=1500)
we have to find another solution.

In Sect. \ref{sec:strongsignal} we considered the extreme case where noise is negligible 
and sky coverage is complete. 
Consider now the opposite limiting case, where the signal is weak compared to noise,
\begin{equation}
S^{-1}\gg A^{\dagger}N^{-1}A .
\end{equation}
Under this assumption we can approximate
\begin{equation}
    (S^{-1}+A^{\dagger}N^{-1}A)^{-1}  \approx S -SA^{\dagger}N^{-1}AS.  \label{Ssmall}
\end{equation}
Matrix $M$ of Eq. (\ref{Mdef}) then simplifies into
\begin{equation}
  M \approx A^{\dagger}N^{-1}A . \label{lowell}
\end{equation}

An explicit formula for $A^\dagger N^{-1}A$ for one detector 
is given by \cite{keihanen2012} as
\begin{equation}
\begin{split}
 & (A^\dagger N^{-1}A)_{s\ell m,s'\ell'm'} =  \\
  & \frac{1}{\sigma^2}
 \sum_{kk'} (-1)^{m'+k'}b_{s\ell k} b^\ast_{s'\ell'k'}    \sum_{\ell_2} (2\ell_2+1) 
          W^{\ell_2}_{m'-m,k'-k} \\
  &     \times
        \left( \begin{array}{ccc}
            \ell & \ell' & \ell_2 \\ m & -m' & (m'\!-\!m)
         \end{array} \right)
          \left( \begin{array}{ccc}
             \ell & \ell' & \ell_2 \\ k & -k' & (k'\!-\!k)\end{array} \right)
         .  \label{artdeco} 
\end{split}
\end{equation}
Here $b_{s\ell k}$ is the harmonic beam expansion,
$\sigma^2$ is the white noise variance in TOI domain,
and $W$ is the Wigner transform of the pointing distribution.
If several detectors are involved,
Eq. (\ref{artdeco}) is replaced by the sum
over all detectors.

The Wigner transform $W$ is computed as an expansion in Wigner functions
of the 3D map data structure, which is the input format assumed
by the \artdeco\ deconvolver.  
A 3D map is a three-dimensional data object constructed from a time-ordered data stream
through a binning operation which combines data samples with similar pointing.
Two of the dimensions are equivalent to the $\theta,\phi$ angles that define an ordinary HEALPix map,
the third keeps track of the distribution of beam orientations.
The Wigner functions offer a natural orthogonal set of base functions 
for this data structure.

When we insert Eq. (\ref{artdeco}) to the kernel formula (\ref{Lambda}),
we obtain a formula with four 3j symbols.
We disconnect $W$ from indices $m,m'$ by writing
\begin{equation}
\begin{split}
   W^{\ell_2}_{m'-m,k'-k} 
   \left( \begin{array}{ccc}
            \ell & \ell' & \ell_2 \\ m & -m' & m'\!-\!m
         \end{array} \right)  =  \\
   \sum_{m_2}     W^{\ell_2}_{m_2,k'-k} 
   \left( \begin{array}{ccc}
            \ell & \ell' & \ell_2 \\ m & -m' & m_2
         \end{array} \right) .
\end{split}
\end{equation}
All dependence on indices $m,m'$
is now in the 3j symbols.
The sums over $m,m'$ can be carried out with the help of the relation
\begin{equation}
 \sum_{mm'} \left( \begin{array}{ccc}
               \ell & \ell' & \ell_2 \\ m & m' & m_2
            \end{array} \right)
            \left( \begin{array}{ccc}
               \ell & \ell' & \ell_3 \\ m & m' & m_3
            \end{array} \right)
    = (2\ell_2+1)^{-1} \delta_{\ell_2\ell_3} \delta_{m_2m_3}.  \label{normalization}
\end{equation}
All this put together, we obtain for the kernel matrix
\begin{equation}
\begin{split}
& \Lambda_{u\ell,u\ell'} =  \\
&  \sum_{\alpha\beta}
\frac{1}{\sigma_{\alpha}^{2}}\frac{1}{\sigma_{\beta}^{2}}
    \sum_{k_1k_2k_3k_4} 
        \sum_{s_1s_2} g^{s_1s_2}_u   
                      b^{1\ast}_{s_1\ell k_1}  b^{2}_{s_2\ell k_2} 
        \sum_{s_3s_4} g^{s_3s_4}_{u'} 
                     b^{1}_{s_3\ell'k_3}  b^{2\ast}_{s_4\ell'k_4}  \\
&   \times
        \sum_{\ell_2}  (-1)^{k_3+k_4}
           \left( \begin{array}{ccc}
             \ell & \ell' & \ell_2 \\ k_1 & -k_3 & k_3-k_1  
                   \end{array} \right) 
           \left( \begin{array}{ccc}
             \ell & \ell' & \ell_2 \\ k_2 & -k_4 & k_4-k_2
                \end{array} \right)              \\
 &  \times  (2\ell_2+1) \sum_{m_2}
            W^{1\ell_2\ast}_{m_2,k_3-k_1} W^{2\ell_2}_{m_2,k_4-k_2}  , \label{BB2} 
\end{split}
\end{equation}
where $\alpha,\beta$ label the detectors involved in the deconvolution process.
This kernel is much cheaper to evaluate than the full kernel which involves inverting
the deconvolution matrix.
The required inputs are the beam expansion $b_{s\ell k}$
and the pointing Wigner transform $W$. 
The latter we can extract from the \artdeco\ code.

Consider then the raw spectrum of Eq. (\ref{Praw}). Under the same approximation 
of Eq. (\ref{Ssmall}) it simplifies into
\begin{equation}
P_{L}  =  \ve t^TN^{-1}A^{\dagger} G^L A N^{-1} \ve t  .
\end{equation}
This is equivalent to taking the spectrum of the right-hand-side of the deconvolution equation.
Note that the prior has disappeared both in the raw spectrum and in the kernel.
The solution thus does not depend on  {\it a priori} information of the CMB spectrum,
as long as we can assume that the CMB signal is weak in comparison with noise.

\subsection{Uniformization}

The obvious problem with the recipe above is that for experiments like \Planck,
the assumption of weak signal does not hold.  
In fact, an attempt to apply the method presented above as it is, 
leads to a gross misestimation of all the spectra.
We can, however,  improve the method through preprocessing of the data.
A clue is found in \cite{hivon2002}, where instead of the actual hit distribution,
a uniform one is assumed when weighting the input map.  It turns out that uniformization of the 
hit count distribution is an essential step in our method too.

The error in a quadratic power spectrum estimation method comes from two sources.
One is the instrument noise, another the statistical variation
of the harmonic coefficients. In the process of deriving the optimal method,
we made the simplification where we replaced the products of the form $a_{s\ell m}a_{s'\ell'm'}$
by their ensemble average. The actual products differ from the ensemble average.
This deviation acts as ``noise'' in the spectrum estimation.
The DQML method finds the optimal balance between the error sources,
and minimizes their combined effect.
The approximation of Eq. (\ref{BB2}), instead,  ignores the CMB-induced error component.
Thus individual strong T multipoles may leak into the regime of much weaker signal,
destroying the estimate there. To prevent this, we would like to minimize the connection
between distant multipoles. 

Analysis of the deconvolution matrix reveals that the correlation between distant multipoles
is related to the uneven distribution of measurements due to scanning strategy. 
This motivates the following processing step.  We {\em uniformize} the sample distribution
by setting the total hit count for each HEALPix pixel that is observed at all, to a common value
which is taken to be the average hit count. This is done by rescaling of the input 3D map object.
In other words, we are reducing the weight of frequently scanned sky pixels.
The distribution of beam orientations
within a pixel is left intact. Further, when working with \Planck\ simulations
we set the assumed noise levels for a detector pair sharing a horn
to the same value. This works further into the direction of reducing the leakage between
temperature and polarization.
We recompute the Wigner transform $W$ with the uniformized distribution as input.
This way, as it turns out, we are able to estimate both the temperature and polarization spectra 
over the full multipole range with very good accuracy 

There is a remarkable similarity between the kernel Eq. (\ref{BB2}) and 
the kernel involved in the {\tt Master} method  \citep{hivon2002}.
An immediate difference is the presence of $k\ne0$ beam coefficients,
and related 3j symbols in our solution.
If we assume perfect delta beams, $b_{0\ell k}=\sqrt{2\ell+1}\delta_{k0}$,
the only non-zero terms of Eq. (\ref{BB2}) are those involving $W^{\ell}_{m0}$,
which is equivalent to the usual harmonic expansion of the hit count.
Under this assumption our result becomes identical to that of {\tt Master}.

In the case of polarization, our kernel can be compared to the corresponding one 
presented by   \cite{kogut2003} and \cite{grain2009}.  
Despite of obvious similarities, here it is more difficult to see how the methods relate,
since the map-based methods involve dividing the polarization signal into Q and U components,
something that does not exist in our method.
What can be said is that assuming perfect delta beams alone is not sufficient to make the results identical.

%%%%%%%%%%%%%%%%%%%%%%%%%%%%%%%%%%%%%%%

\section{Results}

\subsection{Simulations}

We have validated both deconvolution methods with simulations.
The simulation procedure was described in Sect. \ref{sec:simulations}.
As prior in the DQML algorithm we use a CMB spectrum generated from slightly 
different cosmological parameters than those used for the input spectrum.  
The $TT$ prior depicted in Fig. \ref{fig:prior}.

\begin{figure*}
\includegraphics[width=18cm]{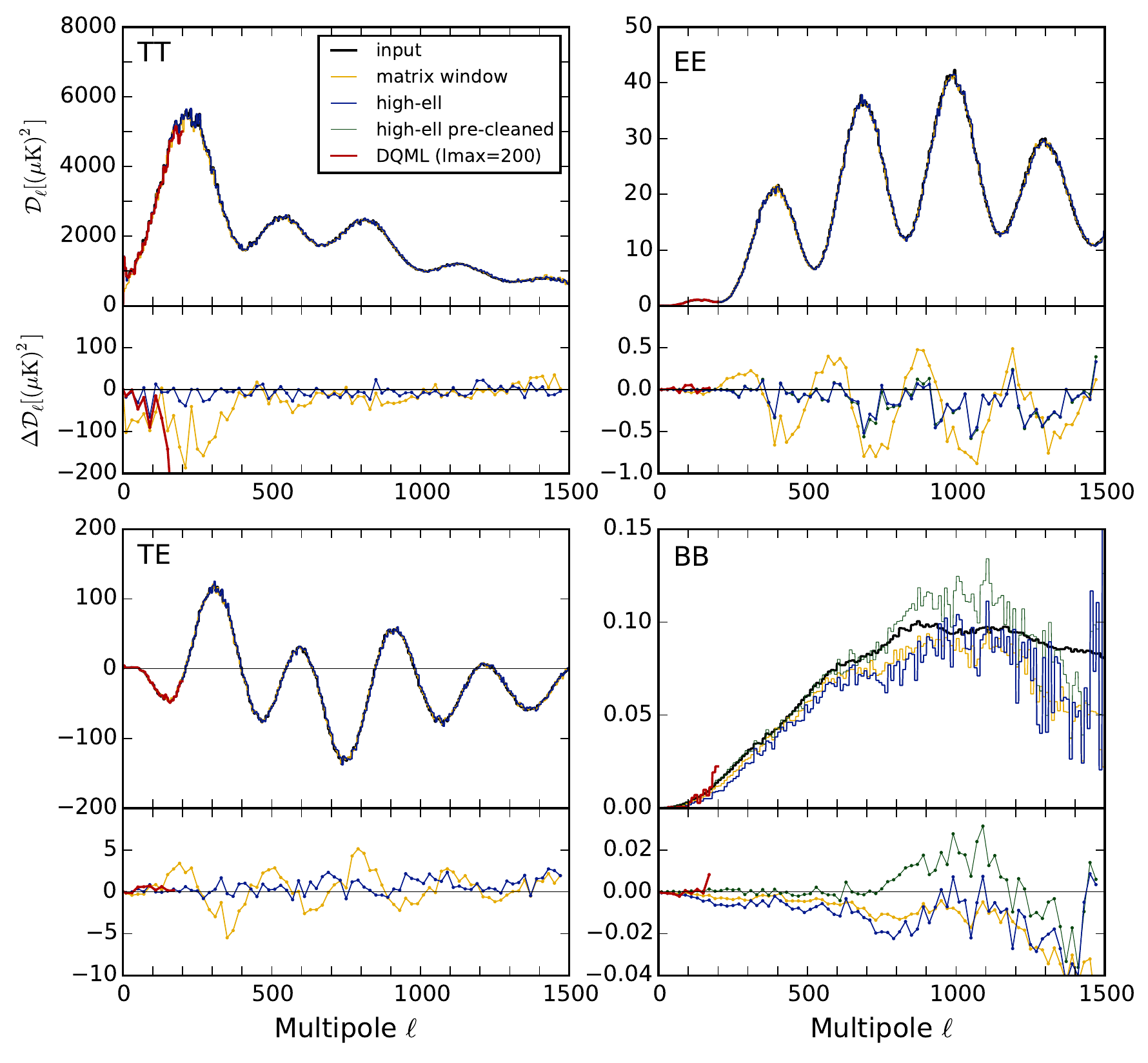}
\caption{
Recovered 70 GHz CMB spectra.
We show the spectrum of the input realization (black)
along with recovered estimates with 
the DQML method up to \lmax=200, 
and the high-ell deconvolution method with \lmax=1500.
For comparison we show also the estimate from the matrix window method 
(best estimate from Fig. \ref{fig:corrections}).
The spectra are averaged over 5 adjacent multipoles (10 for $BB$), the error over 20 multipoles.
Note the tighter scale of the error plot as compared to Fig. \ref{fig:corrections}.
We show also the effect from pre-cleaning of temperature signal (only visible in $EE$ and $BB$).
}
\label{fig:cmb_70GHz_kernelcorr}
\end{figure*}

The main results for the fiducial 70 GHz simulation are shown in Fig. \ref{fig:cmb_70GHz_kernelcorr}.
We plot in the same figure spectrum estimates from the DQML method of Sect. \ref{sec:optimal}
and the high-ell method of Sect. \ref{sec:noisedom}. 
We apply the DQML method to multipoles $\ell=0-200$,
the high-ell method to multipoles $\ell=0-1500$.
For comparison we show also results from the {\tt PolSpice} mask correction combined with 
matrix window correction,
which was the best of the pixel-based methods
of Sect. \ref{sec:pixelmethods}. 
We refer to this combined method for brevity as ''matrix window`` method.

All three methods recover the CMB spectrum well in their applicable multipole range.
An exception is the $BB$ spectrum where the high-multipole methods show a small negative bias.
Differences between the methods become visible in the lower panel, where we plot
the absolute error (difference between the recovered spectrum and the input spectrum).
We observe that in all spectra but $BB$, the high-ell deconvolution method yields a still more accurate estimate than
the matrix window method. The improvement is particularly clear in $TT$ around the first acoustic peak.
Also, the matrix window method for  $EE$ and $TE$ shows residuals of temperature leakage,
which are reduced by the high-ell deconvolution method.
In $BB$, matrix window and high-ell deconvolution give similar results, 
unless we apply pre-cleaning (see Sect. \ref{sec:precleaning}).

The drop in the DQML estimate for $TT$ above $\ell=100$ is a fringe effect,
caused by the limited \lmax value. Extending the multipole range across the first acoustic peak
would probably reduce the effect.

\begin{figure*}
\includegraphics[width=18cm]{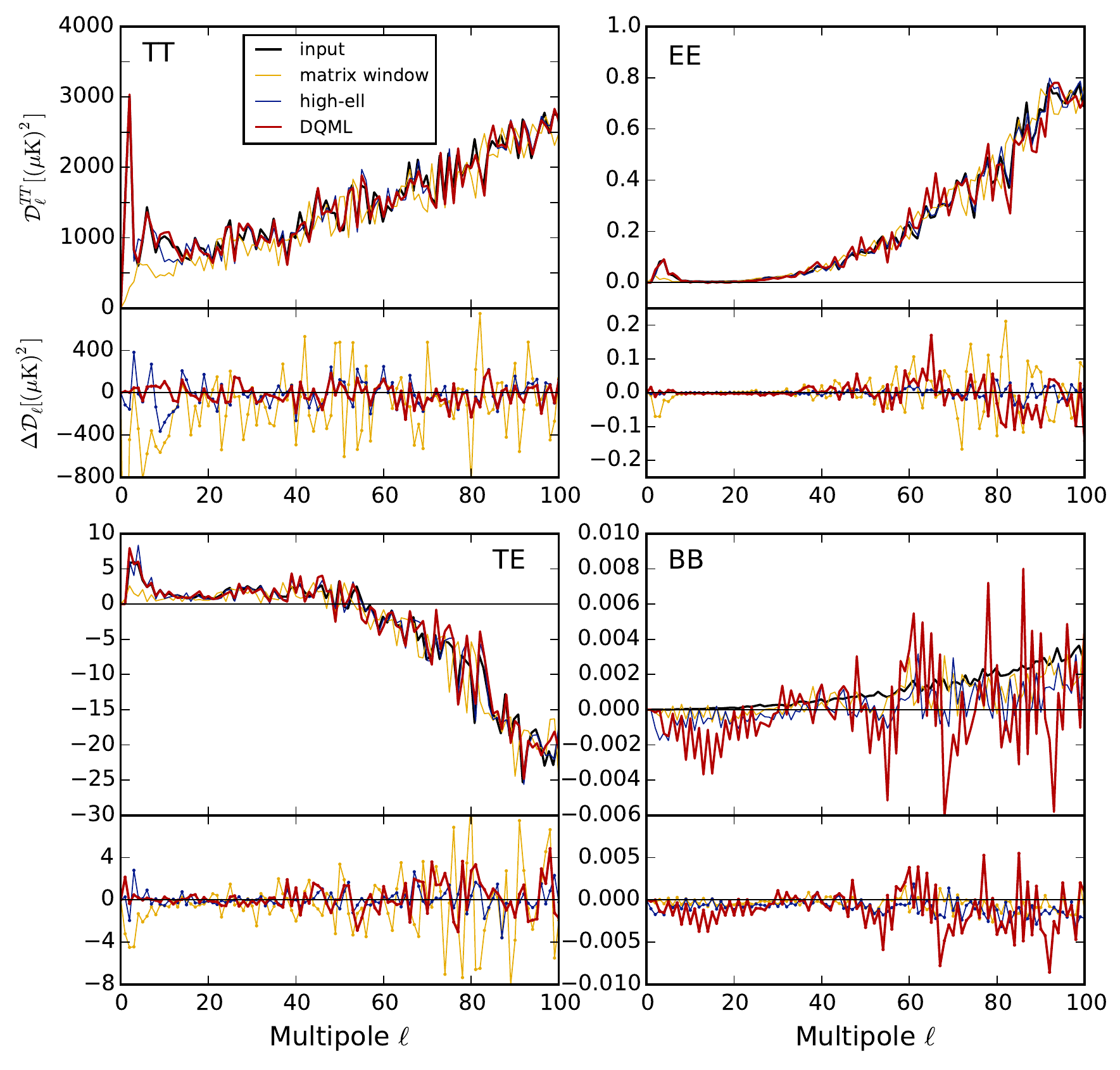}
\caption{
Zoom into the low multipole region.
Data and line types are the same as in Fig. \ref{fig:cmb_70GHz_kernelcorr}.
We show all the individual multipoles without binning.
}
\label{fig:zoom}
\end{figure*}

We blow up the low-multipole region in Fig \ref{fig:zoom}.
Here we show the individual multipoles without binning.
Consequently, the amplitude of error is larger than in Fig. \ref{fig:cmb_70GHz_kernelcorr}.
The high-ell method is again superior to the matrix window method,
and in $EE$ and and in $BB$ it is also more accurate than the DQML method.
At the very lowest $TT$ multipoles ($\ell<40$) the DQML method gives the most accurate estimate.
The differences, however, are at the same level as the cosmic variance.
We see also that the two deconvolution methods give nearly identical results in $TT$
at  intermediate multipoles ($20<\ell<100$).  This means that the two estimates
can be joined without discontinuity.
The $TE$ spectrum behaves similarly to $TT$, i.e. DQML is more accurate at lowest multipoles,
 but the differences are less clear than in $TT$

%%%%%%  FIGURES %%%%%%%%%%%%%%

%%%%%%%%%%%%%%%%%%%%%%%%%%%%%%%%%

\subsection{Pre-cleaning}
\label{sec:precleaning}

As discussed in Sect. \ref{sec:noisedom}, the strong low-ell T signal leaking to higher multipoles or to polarization
interferes with the estimation at these weaker spectral components.  
The procedure of uniformizing the hit count distribution
was introduced to mitigate the problem. We can do more than that, as follows.
We run the usual deconvolution procedure (with prior) to obtain an estimate of the \aslm\ coefficients of the sky.
We then scan the T components back into a TOI stream, subtract it from the
original TOI, and perform the usual power spectrum estimation procedure on the cleaned data,
to obtain a better estimate of the weaker spectrum components.
The signal components to be subtracted can be chosen in many ways.
In Fig \ref{fig:cmb_70GHz_kernelcorr} we show the effect on $EE$ and $BB$ of removing all of the T signal.
In $EE$ the effect is negligible, but in $BB$ pre-cleaning step removes the negative bias almost
perfectly up to $\ell=800$.

\begin{figure}
\includegraphics[width=8.8cm]{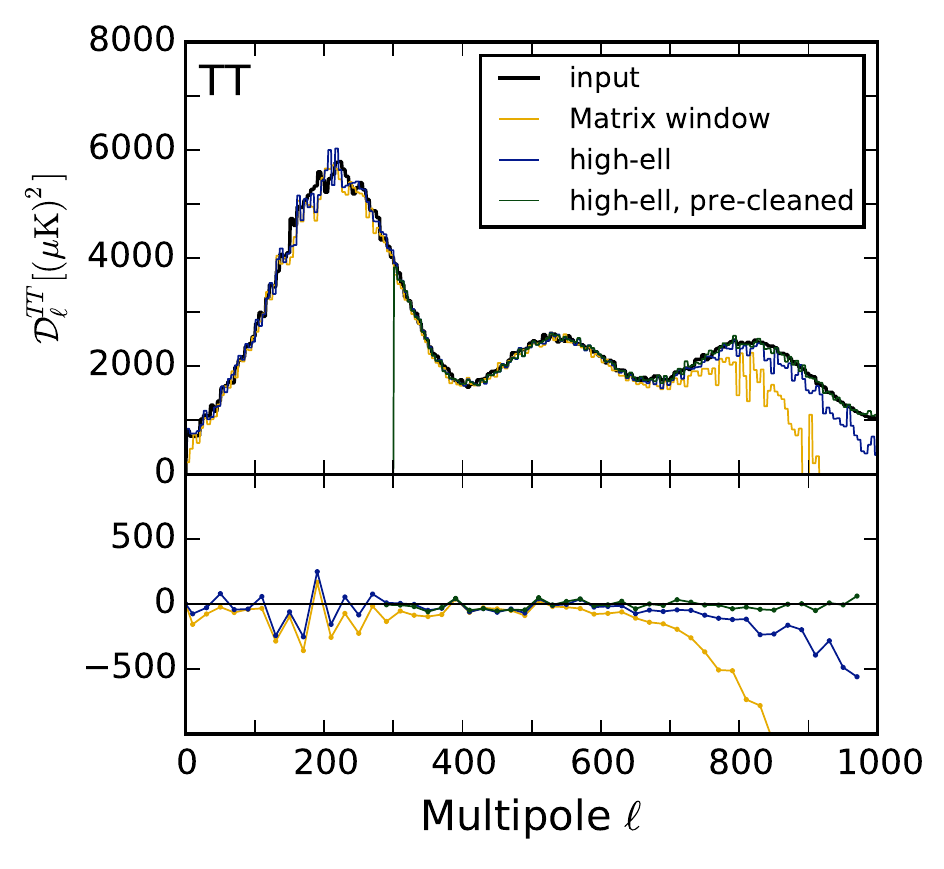}
\caption{
An example of a case where pre-cleaning is important. 
Shown is the recovered 44 GHz $TT$ spectrum compared against input.
The strong quadrupole moment leaks into the high multipole regime
in a way that is not captured by the spectrum estimation methods.
Subtracting an estimate of the low multipole signal up to $\ell$=300 removes the bias.
}
\label{fig:cmb_44GHz_kernelcorr}
\end{figure}

In the case above, pre-cleaning has little practical meaning, since the small bias in the $BB$ spectrum 
is well below the noise level of any current CMB experiment.
In some cases, however, pre-cleaning may become important.
One such case is shown in Fig \ref{fig:cmb_44GHz_kernelcorr}. 
We show the recovered $TT$ spectrum from a simulation involving the 44 GHz channel of \Planck.
Beam leakage effects are known to be particularly strong for this channel.
Without pre-cleaning, the spectrum is underestimated at high multipoles ($\ell<700$).
Note that this is not specific to deconvolution,
but the matrix window method suffers from the same problem.
Analysis of individual multipoles revealed that the bias arises from the very lowest temperature
multipoles, which leak into the high-multipole regime. 
We pre-cleaned the 44 GHz data set by subtracting the deconvolved T multipoles up to \lmax=300.
This removed the bias entirely.  
This particular realization happened to have a very strong quadrupole,
which was found to be responsible for most of the leakage.
The quadrupole of the actual CMB sky, instead, is known to be small. The simulation shown here probably
represents an overly pessimistic case.  However, leakage effects depend on the specifics beam shapes
and the scanning strategy of a particular experiment, and it is difficult to predict where such effects may arise.
It is thus advisable to check the robustness of the results against pre-cleaning parts of the data,
or else to validate the results with simulations.

%%%%%%%%%%%%%%%%%%%%%%%%%%%%%%%%%%%%

%%%%%%%%%%%%%%%%%%%%%%%%%%%%%%%%%%

\subsection{Instrument noise}

At first it seems surprising that the optimal method appears to be inferior to the high-ell method 
for all but the lowest $TT$ multipoles, 
considering that the optimal method was specifically designed to minimize the error. 
There is an explanation for this.
When deriving the optimal method we required that it minimizes the {\em combined} error arising from 
statistical variation in the CMB itself, 
and from instrument noise.
So far we have applied the method to a pure CMB simulation.
It is conceivable that when instrument noise is taken into account,
the optimal method will give a smaller total error.

\begin{figure}
\includegraphics[width=8.8cm]{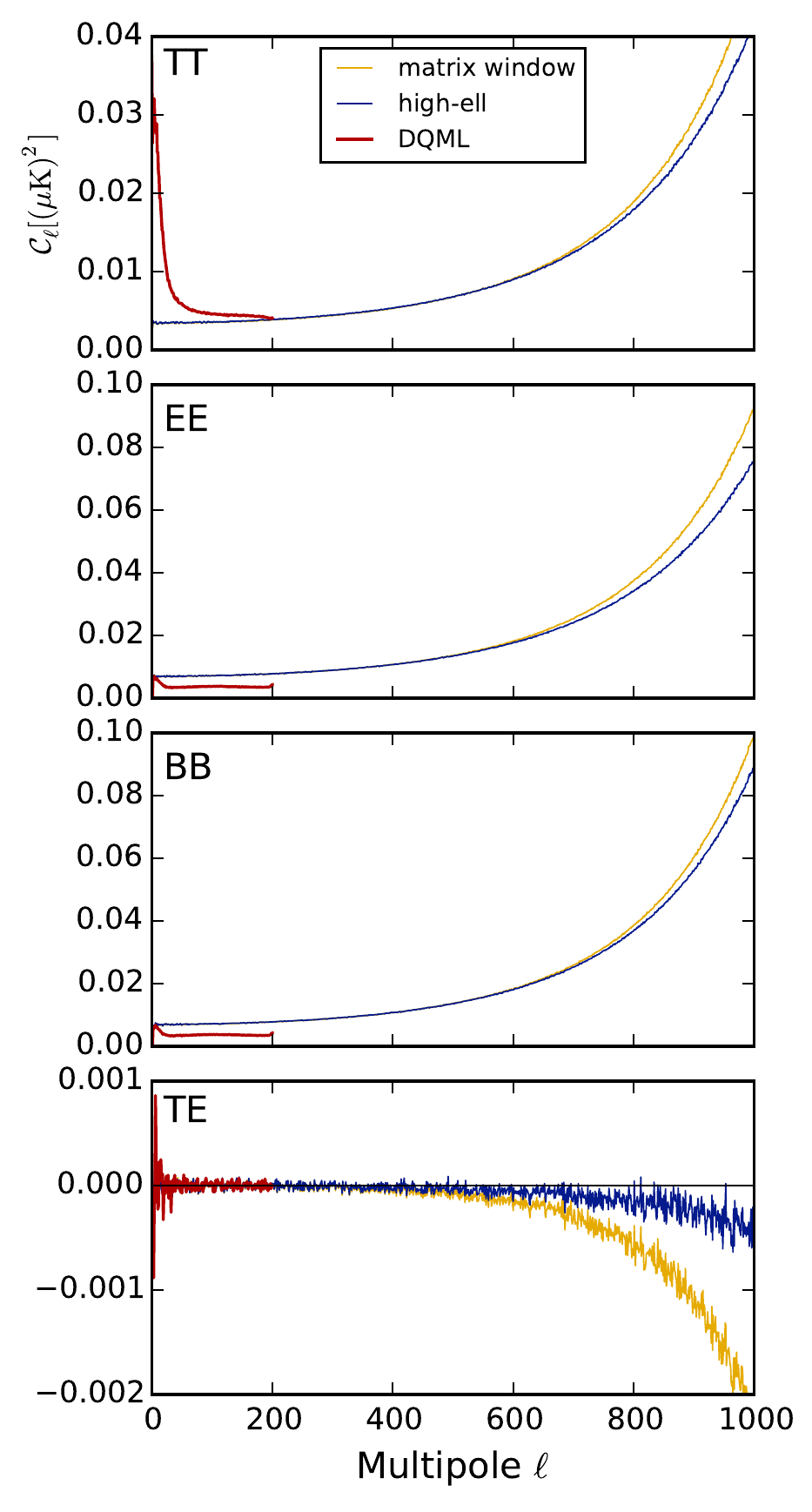}
\caption{
Noise bias,
assuming white noise level of \Planck\ 70 GHz detectors.
as estimated from Monte Carlo simulations.
The noise bias is plotted in the ${\cal C}_\ell$ convention,
i.e. without scaling by $\ell(\ell+1)/(2\pi)$.
}
\label{fig:noisebias}
\end{figure}

To verify if this is the case, we perform a series of noise simulations.
We generate 100 realizations of white noise,
at the level of \Planck\ 70 GHz channel.  
The parameters were taken from \cite{planck2014-a03}.
We processed the noise through the same power spectrum estimation 
pipeline as the signal simulations,
with the three nethods in question.
For each method we obtain 100 noise spectra with realistic properties.
We compute the {\em noise bias} as the average over the 100 noise spectra.
The noise bias for the three methods are plotted in Fig. \ref{fig:noisebias}.  
We plot the spectrum without the $\ell(\ell+1)/(2\pi)$ scaling,
to avoid excessive amplification at high multipoles.
We cut the plot at $\ell$=1000 to show the low multipole region more clearly.

The noise bias for the matrix window method and for the high-ell deconvolution method
are very close at low multipoles.  At high multipoles ($\ell>600$)
the bias for the matrix window method begins to increase more steeply.
The noise bias for the DQML method behaves very differently.

The noise bias itself is not a good measure of residual noise.
The noise bias can be estimated through Monte Carlo simulations,
and subtracted from the actual spectrum.
What is more interesting is the variation of a single realization,
that remains after the subtraction of the mean noise bias.
This is the residual noise in the final spectrum estimate.
We estimated the residual noise level from the same MC simulations that we used for 
the estimation of the noise bias.
We subtracted the mean bias from the 100 noise spectra, sorted the 100 values at each individual multipole, 
and picked the 16th and 85th value.
These values correspond to one-sigma variation
in both directions around the mean noise bias.

\begin{figure}
\includegraphics[width=8.8cm]{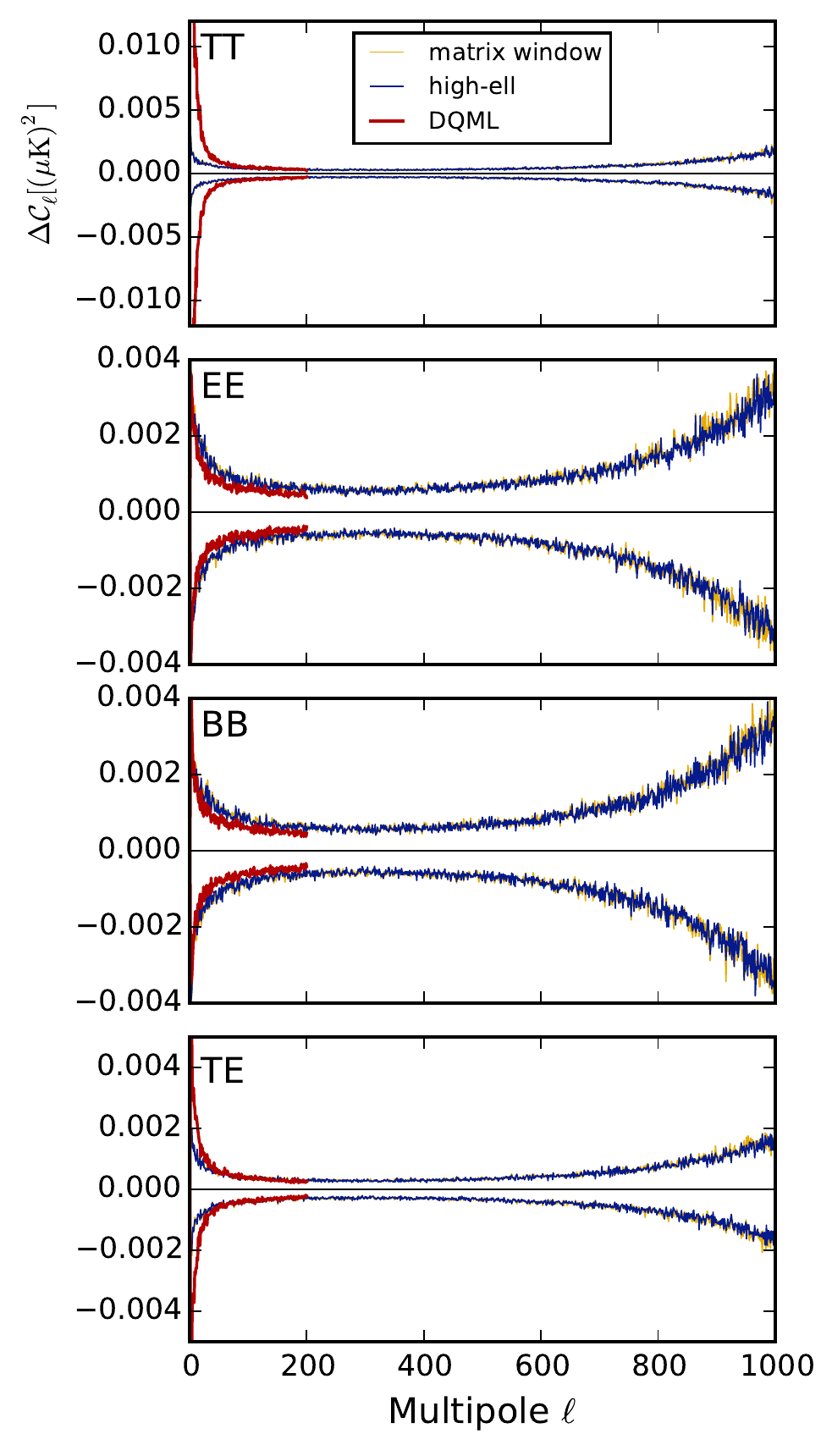}
\caption{
Residual noise level in the recovered spectra,
after subtraction of the average noise bias.
We assume the white noise level of \Planck\ 70 GHz detectors.
Shown is the one-sigma variation of the residual noise in both directions
in ${\cal C}_\ell$ units,
as estimated from 100 Monte Carlo realizations.
}
\label{fig:noisevariation}
\end{figure}

The results are shown in Fig. \ref{fig:noisevariation}.  
The DQML method yields a higher residual noise than the high-ell method in the $TT$ spectrum.
The noise level is, however, negligible compared with the signal-induced error of Fig. \ref{fig:zoom},
where DQML performed better.
In $EE$ and $BB$ components, the DQML method gives a significantly lower residual noise level
than the high-ell method.
Since noise is the dominant error source in these spectra,
this is consistent with the theoretical expectation that the DQML method minimizes the total error.

%%%%%%  FIGURES %%%%%%%%%%%%%%

%%%%%%%%%%%%%%%%%%%%%%%%%%%%

\subsection{Cross-power spectrum}

In this work we have made the simplifying assumption that the instrument noise is white.
This not exactly true for realistic instruments. The TOI of \Planck, for instance,
is contaminated by slowly-varying $1/f$ noise.  The correlated noise component 
can partly be removed by destriping techniques, but there remains a residual
that contaminates the lowest multipoles.   If noise properties are well known,
the noise bias can be estimated through MC simulations, and subtracted from
the spectrum estimate, to yield an unbiased spectrum estimate.

Another possibility is to cross-correlate two independent data sets.
If noise is uncorrelated from one set to another, the noise bias vanishes,
and the estimate is again unbiased.
Both deconvolution PSE methods presented in this paper 
extend trivially to cross-power estimation.
Eq. (\ref{Praw}) is replaced by
\begin{equation}
  P_{L}  =  {\hat {\ve a_1}}^{\dagger} S^{-1} G^{L\dagger} S^{-1} \hat {\ve a_2}  ,  \label{Praw_cross}
\end{equation}
where index $1,2$ refers to the two data sets.
The kernel of Eq. (\ref{Lambda}) takes the form
\begin{equation}
  \Lambda_{LL'} = \mathrm{Tr}\left[M_1 G^{L}M_2^\dagger G^{\dagger L'}\right] . \label{Lambda_cross}
\end{equation}
The high-ell kernel of Eq. (\ref{BB2}) is already written for a detector pair,
and requires no modification.

\begin{figure}
\includegraphics[width=8.8cm]{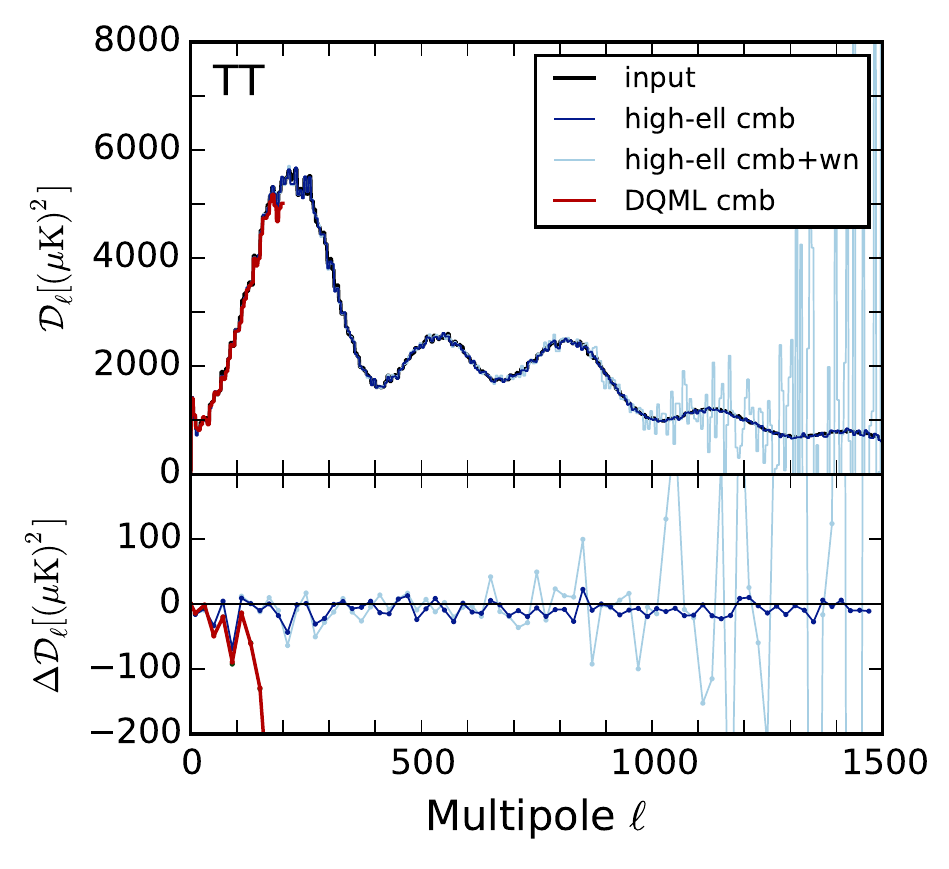}
\caption{
Cross-power spectrum estimation. Shown is the 70 GHz $TT$ cross-spectrum between two two-year data sets,
for pure CMB (dark blue), and for one realization of white noise at \Planck\ level.
The fiducial 4 year data set is split into two halves, which are
cross-correlated to give an unbiased spectrum estimate.
}
\label{fig:yr12}
\end{figure}

We demonstrate the cross-power estimation in Fig. \ref{fig:yr12}.
We split the fiducial 70 GHz simulation into two halves, one consisting of years 1-2 and the other of years 3-4.
The cross-power $TT$ estimate is remarkably similar to the auto-spectrum estimate of 
Fig. \ref{fig:cmb_70GHz_kernelcorr}.
The same lows and bumps are present in both error plots,
indicating that the error arises from the CMB sky itself.

We plot also the cross-power spectrum from a simulation where
we add one realization of white noise at \Planck\  level.
In the range $\ell<200$ where the DQML method is applied, noise is negligible.
The results from the simulation with noise are practically identical to those from 
the pure CMB simulation, and are not shown separately.

%%%%%%%%%%%%%%%%%%%%%%%%%%%%%%%%

%%%%%%%%%%%%%%%%%%%%%%%%%%%%%%%%%%%%%%%%%%%%%%%

\section{Statistical kernel evaluation}
\label{sec:statistical}

While the light-weight high-ell method works very well in most cases,
our simulations have indicated that the optimal DQML method
leaves a lower residual noise level.
It was also superior in determining the lowest signal-dominated $TT$ multipoles.
With future CMB experiments also the polarization signal may enter the regime of signal domination.
It is thus worth the effort to explore ways of extending the method to a wider multipole range. 
The bottleneck is the current implementation is the construction and inversion of the 
 deconvolution matrix. The rank of the matrix grows as proportional to $\lmax^2$,
 and the CPU cost of the inversion as $\lmax^6$.
 It is thus unlikely that the multipole limit can be pushed much further even with increased
 computational resources. A numerical method that allows to evaluate the kernel of Eq. (\ref{Lambda}) 
 without explicit matrix inversion
 would be highly welcome, but such a method is not in sight.

We explore an alternative way of evaluating the kernel matrix,
involving Monte Carlo simulations.
To start, we rewrite matrix $M$ from Eq. (\ref{Mdef}) as
\begin{equation}
M =  A^\dagger N^{-1}A (S^{-1}+A^\dagger N^{-1}A)^{-1} S^{-1} .  \label{Mother}
\end{equation}
Here $N=\langle \ve n \ve n^T \rangle$ is the white noise covariance in TOI domain. 
We rewrite further
\begin{equation}
N^{-1}  = N^{-1}  \langle \ve n \ve n^T  \rangle N^{-1}  , \label{Nsplit}
\end{equation}
where $\ve n$ denotes a white noise realization.
We insert this into the first occurrence of $N$ of Eq. (\ref{Mother})
and that further into Eq. (\ref{Lambda}).
Since $M$ appears there twice, we must introduce two independent noise 
realizations $\ve n_1$ and $\ve n_2$,
so that we can combine the averages as
\begin{equation}
\langle \ve n_1\ve n_1^T \rangle \langle \ve n_2\ve n_2^T \rangle 
=\langle \ve n_1\ve n_1^T \ve n_2\ve n_2^T \rangle .
\end{equation}
Because a permutation operation leaves the trace unchanged, we can rearrange
the terms to obtain
\begin{equation}
\begin{split}
&\Lambda = \langle \mathrm {Tr}[ \ve n_1^T N^{-1}A G^L A^\dagger N^{-1} \ve n_2
\ve n_2^T N^{-1}A(S^{-1}+A^\dagger N^{-1}A)^{-1} S^{-1}  \\
& G^{L'\dagger} S^{-1}(S^{-1}+A^\dagger N^{-1}A)^{-1} A^\dagger N^{-1}\ve n_2
] \rangle . \label{statkernel}
\end{split}
\end{equation}
This lengthy construction actually consists of two scalar terms 
of the form $\ve n_1^T\cdots\ve n_2$.
We can thus drop the trace operation (the trace of a scalar is the scalar itself).
The kernel as given by Eq. (\ref{statkernel})
 can be evaluated through MC simulations as follows. We first generate a sequence of 
 white noise realizations $\ve n_k$, $k=1\ldots n_{\mathrm mc}$.  
 For each realization we evaluate the objects
 \begin{eqnarray}
 \ve a_k &=& (S^{-1}+A^\dagger N^{-1}A)^{-1} A^\dagger N^{-1}\ve n_k \nonumber \\
 \ve r_k &=& A^\dagger N^{-1} \ve n_k  \label{ark} .
 \end{eqnarray}
Both objects have the structure of an \aslm\ vector.
The first line represents the usual deconvolution operation.
The second operation is simply the right-hand-side of the deconvolution equation.
Both products are standard outputs of \artdeco\ code.
We thus already have all the necessary machinery in place.
We evaluate the spectra
\begin{eqnarray}
P^L_r &=& \ve r_j^T G^L \ve r_k \nonumber \\
P^L_a &=& \ve a_j^T S^{-1} G^LS^{-1} \ve a_k
\end{eqnarray}
for all realization pairs $j\ne k$.
This is equivalent to the familiar operation of evaluating the cross-spectrum
between two harmonic vectors, apart from normalization.
An estimate for the kernel is obtained as an average over all pairs
\begin{equation}
\hat \Lambda_{LL'} = \frac{1}{n_{\mathrm mc}(n_{\mathrm mc}-1)}\sum_{j\ne k} 
\ve r_j^\dagger G^L \ve r_k \cdot \ve a_k^\dagger S^{-1} G^{L'} S^{-1} \ve a_j  ,\label{Lambdastat}
\end{equation}
where $n_{\mathrm mc}(n_{\mathrm mc}-1)$ 
counts the pairs $j\ne k$.
The estimate converges to the true kernel when $n_{\mathrm mc}\rightarrow\infty$.
Note that the noise simulations here have nothing to do with the actual noise properties
 of the data. We are simply evaluating the diagonal matrix $N$ statistically.

\begin{figure}
\includegraphics[width=8.8cm]{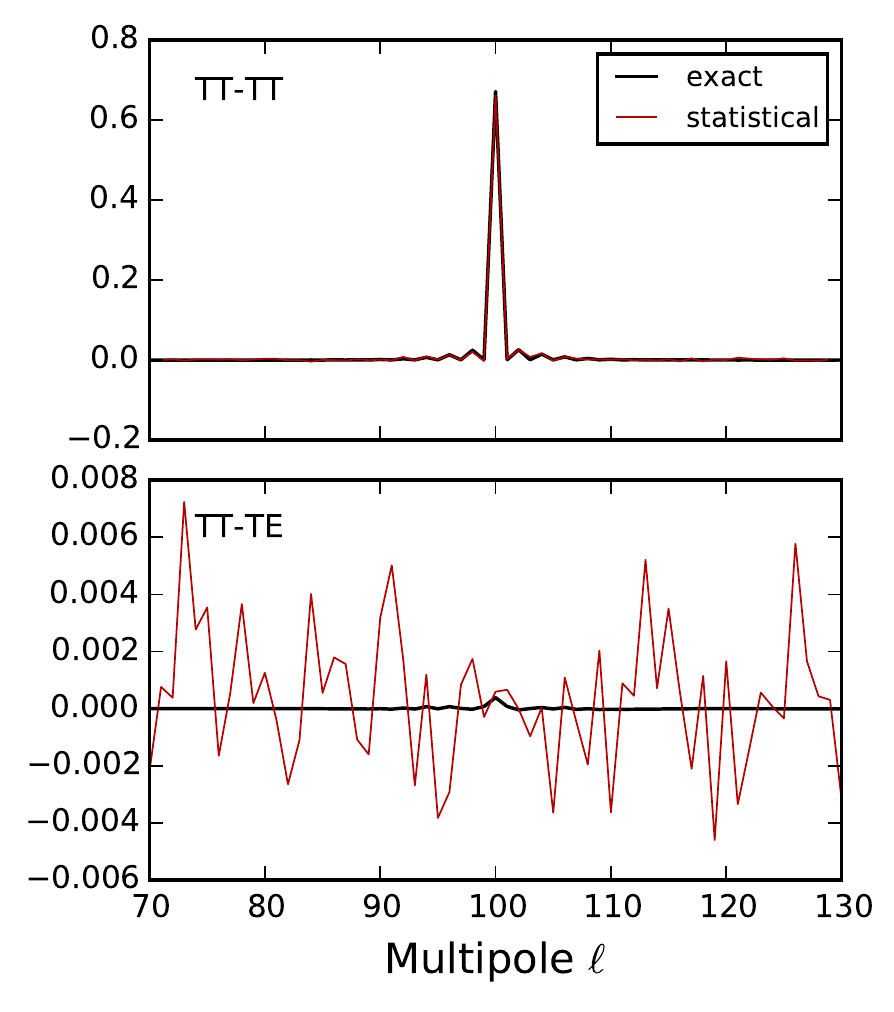}
\caption{
Selected elements of the 30 GHz kernel, evaluated through the exact numerical 
method of Sect. \ref{sec:optimal} and the statistical method of Sect. \ref{sec:statistical}.
Shown are elements along the row corresponding 
to multipole $\ell$=100 in $TT$.  The $TT$-$TT$ elements are recovered well with the statistical method,
but the number of realizations (800) is insufficient for the weaker $TT$-$TE$ elements.
}
\label{fig:matrix_elements}
\end{figure}

To validate the procedure we pick the less demanding Planck 30 GHz channel,
and generate 800 white noise realizations.  
The procedure took roughly one minute of wall-clock time per realization on 768 cores.
We evaluate the kernel matrix with the statistical method up to \lmax=800,
and with exact DQML up to \lmax=200, and compare the kernel elements in the multipole 
range covered by both methods.  We plot elements of row $\ell$=100 in Fig. \ref{fig:matrix_elements}.
The 800 realizations we have are sufficient to recover the structure of the $TT-TT$ block of the kernel,
but the weaker $TT-TE$ cross terms are totally buried under statistical variation.
We conclude that a lot more realizations are required,
for the method to become useful in practice.

%%%%%%%%%%%%%%%%%%%%%%%%%%%%%%%%%%%%%%

%%%%%%%%%%%%%%%%%%%%%%%%%%%%%%%%%%%%%%

\section{Conclusions}

We have studied the possibility of combining beam deconvolution technique with power spectrum 
estimation (PSE) for absolute CMB measurements.  We present two new power spectrum estimation methods,
one suitable for low multipoles, the other for a wide multipole range.
The methods yield an estimate for the angular CMB power spectrum in both temperature and polarization.
The methods correct simultaneously the effect of incomplete sky coverage
and of asymmetric beam shape. In particular, they correct the leakage from temperature
to polarization through beam shape mismatch.
The required inputs are the time-ordered data, pointing information, and known beam shapes.  

The first method (DQML) was
derived formally as the quadratic estimator that fulfils two requirements: 
it yields a non-biased estimate of the CMB spectrum, and at the same time 
minimizes the residual error in the estimate.
The residual error is a combination of instrument noise, and the error that arises from the 
statistical variation of the individual \aslm\ sky coefficients.

We developed the optimal solution into a practical algorithm. 
The procedure consists of beam-deconvolving the data with a first-guess CMB spectrum as prior,
and correcting the spectrum with a kernel that 
takes into account the exact pointing distribution and beam shapes.
The computational burden grows steeply with increasing \lmax.
We successfully applied it to the multipoles in range $\ell$=0--200.

For the high multipole regime we derived another solution, which is suboptimal in the sense
 that it does not minimize the residual error, but still yields an unbiased spectrum estimate.
 Formally it is derived as the weak-signal limit of the optimal DQML method.
The high-ell method is computationally light, and 
we apply it to multipoles up to $\ell$=1500.  
The core of the method is a kernel matrix (Eq. (\ref{BB2})),
which simultaneously corrects for the sky coverage and for beam effects.
The kernel resembles the one implemented in the {\tt Master} method,
but includes additional terms which take into account the beam.

We validated the methods with simulated data. Our fiducial data set mimics that of the \Planck\ LFI 70 GHz channel.
We used realistic \Planck\ beams and scanning strategy.
We applied a galactic mask with $f_\mathrm{sky}$=0.8967 sky coverage.
We found the DQML method to be superior
when estimating the $TT$ spectrum in the low multipole range ($\ell<40$).  

The high-ell method appears superior to the DQML method in the polarization components,
when we deal with pure CMB simulations.
When we add white noise at \Planck\ level, however, we observe that DQML
produces a lower residual noise level.

The best estimate for the CMB spectrum is obtained by a combination of the two deconvolution methods.
For low multipoles it makes sense to use the DQML method, for high multipoles the high-ell method.
The two estimates overlap cleanly at intermediate multipoles with no apparent discontinuity.

We compared the new methods with selected pixel-based PSE methods.
The most accurate of those was one that combines {\tt PolSpice}
mask correction with a matrix beam window function.  
The new methods were found to be still more accurate in the studied simulation case.
Apart from accuracy, the new methods offer the benefit that
they avoid the need for heavy CMB Monte Carlo simulations
which are usually needed in the evaluation of a beam window function.

The new methods operate at TOI level, which is the natural domain for handling 
effects that depend on time or on detector orientation. 
While this work focuses on asymmetric beams, we can consider the possibility 
of extending the general idea to other time-dependent effects.
For instance, a long detector time constant can be included in the beam model.
Further, in the current implementation it is assumed that residual noise at TOI level
is at least approximately white.  The formulation is more general, however,
and could in principle be extended to incorporate correlated noise residuals.
This is achieved by replacing  the diagonal covariance $N$ by an appropriate noise filter.

Beam effects are usually thought to be unimportant at low multipoles.
One may therefore ask if the DQML method provides any benefit over 
the beamless QML method.  This is a valid question, and a definite answer
would require running DQML and {\tt BolPol} or equivalent side by side.
There is, however, an important difference between the methods,
unrelated to beam shapes.  The QML method, as implemented in {\tt BolPol},
takes as input low-resolution CMB maps.
The process of downgrading a high-resolution map is not a trivial one,
and it can be done in different ways (the problem in the context 
of \Planck\ is discussed in \cite{planck2014-a06}).
The DQML method avoids all these issues, as it always operates at the full resolution
of the input 3D map.  The transition to low resolution occurs in a natural way in harmonic space,
where we simply cut the raw spectrum at the desired \lmax\ before applying the kernel.
The DQML method is thus not equivalent to the QML method,
even if we assume perfect delta beams.

Finally we studied the possibility of extending the DQML method to higher multipoles.
We showed that the kernel matrix can be evaluated statistically through
white noise Monte Carlo simulations, but found that 800 noise realizations do not give
sufficient accuracy yet.  Still this is a promising idea to study further,
since the error variance goes down as inversely proportional to
the number of MC realizations.   This is to be contrasted with the analytical method,
where the computational burden grows as proportional to $\lmax^6$ with increasing \lmax.

\section*{Acknowledgements}

This work was supported by the Academy of Finland grants 253204, 257989, 283497, and  295113.
KK is supported by the Magnus Ehrnrooth Foundation.
MR is supported by the German Aeronautics Center and Space Agency (DLR), under
program 50-OP-0901, funded by the Federal Ministry of Economics and
Technology. 
The sky masks and detector beams used in this work are available via Planck Legacy Archive.
 They are based on observations obtained with Planck, an ESA science mission with instruments and contributions 
 directly funded by ESA Member States, NASA, and Canada.
Some of the results in this paper have been derived using the HEALPix
package. 
We thank CSC -- IT Center for Science, Finland -- for computational resources.

\bibliographystyle{mnras}
\bibliography{Planck_Helsinki_bib}

\bsp	% typesetting comment
\label{lastpage}
\end{document}